\documentclass[letterpaper,twoside,10pt]{article}
\usepackage{amssymb}
\usepackage{amsmath}
\usepackage{latexsym}
\usepackage{verbatim}
\usepackage{epsfig}
\usepackage{stmaryrd}
\usepackage{fancyheadings}
\usepackage{pstricks,pst-node}
\usepackage{rotating,graphicx}
\usepackage[greek,english]{babel}
\newlength{\figurewidth}
\newlength{\enviropost}
\newcommand{\be}{\begin{equation}}
\newcommand{\ee}{\end{equation}}
\newcommand{\ble}[1]{\begin{equation} \label{#1}}
\newcommand{\bae}{\begin{eqnarray}}
\newcommand{\eae}{\end{eqnarray}}

\newcommand{\nn}{\nonumber}
\newcommand{\ff}{\nn \\}
\newcommand{\fe}{& = &}

\newcommand{\mvb}{\mathversion{bold}}
\newcommand{\mvn}{\mathversion{normal}}

\newcommand{\ip}[2]{\left\langle #1, #2\right\rangle}

\newcommand{\ket}[1]{| #1 \rangle}

\newcommand{\eg}{\hbox{\em e.g.{}}}
\newcommand{\etc}{\hbox{\em etc.{}}}
\newcommand{\ie}{\hbox{\em i.e.{}}}
\newcommand{\wrt}{\hbox{w.r.t.{}}}
\newcommand{\lhs}{\hbox{l.h.s.{}}}
\newcommand{\rhs}{\hbox{r.h.s.{}}}

\newcommand{\capitem}[1]{\caption{\textsf{#1}}}

\newcommand{\calD}{\mathcal{D}}

\newcommand{\calG}{\mathcal{G}}
\newcommand{\calH}{\mathcal{H}}

\newcommand{\calL}{\mathcal{L}}

\newcommand{\calO}{\mathcal{O}}

\newcommand{\calR}{\mathcal{R}}

\newcommand{\sele}{\selectlanguage{english}}

\newcommand{\selg}{\selectlanguage{greek}}
\newcommand{\Alt}{\operatorname{Alt}}
\newcommand{\eabc}{\epsilon_{ab}^{\phantom{ab}c}}

\newcommand{\GG}{\calG_{\text{G}}}
\newcommand{\GCR}{\calG_{\text{CR}}}
\newcommand{\GP}{\calG_{\text{P}}}
\newcommand{\GPH}{\calG_{\text{PH}}}
\newcommand{\GQR}{\calG_{\text{QR}}}
\newcommand{\muCR}{\mu_{_\text{CR}}}
\newcommand{\muPH}{\mu_{_\text{PH}}}

\newcommand{\psiH}{\psi_{_\text{H}}}
\newcommand{\psiPMZ}{\psi_{_\text{PMZ}}}
\newcommand{\psiZMP}{\psi_{_\text{ZMP}}}
\newcommand{\psiPMP}{\psi_{_\text{PMP}}}
\newcommand{\psiZMZ}{\psi_{_\text{ZMZ}}}
\newcommand{\psiPPJ}{\psi_{_\text{PPJ}}}
\newcommand{\psiZZJ}{\psi_{_\text{ZZJ}}}
\newcommand{\psiPZJ}{\psi_{_\text{PZJ}}}
\newcommand{\phiJJ}{\phi_{_\text{JJ}}}
\newcommand{\phiJK}{\phi_{_\text{JK}}}
\newcommand{\phiKJ}{\phi_{_\text{KJ}}}
\newcommand{\phiKK}{\phi_{_\text{KK}}}

\newcommand{\chiJJJ}{\chi_{_\text{JJJ}}}
\newcommand{\chiJJK}{\chi_{_\text{JJK}}}
\newcommand{\chiJKJ}{\chi_{_\text{JKJ}}}
\newcommand{\chiJKK}{\chi_{_\text{JKK}}}
\newcommand{\chiKKJ}{\chi_{_\text{KKJ}}}
\newcommand{\chiKKK}{\chi_{_\text{KKK}}}
\sele
\newcommand{\papertitle}{%
Generalized Quantum Relativistic Kinematics:\\[2mm] a Stability Point
of View%
}
\newcommand{\runningtitle}{%
Generalized Quantum Relativistic Kinematics: a Stability Point
of View%
}
\newcommand{\paperauthor}{%
C.{} Chryssomalakos and E.{} Okon%
}
\begin{document}
\begin{titlepage}
\vspace*{-1cm}
\begin{flushright}
\textsf{}
\\
\mbox{}
\\
\textsf{20th October 2004}
\\[3cm]
\end{flushright}
\renewcommand{\thefootnote}{\fnsymbol{footnote}}
\begin{LARGE}
\bfseries{\sffamily \papertitle}
\end{LARGE}

\noindent \rule{\textwidth}{.6mm}

\vspace*{1.6cm}

\noindent \begin{large}%
\textsf{\bfseries%
\paperauthor
}
\end{large}


\phantom{XX}
\begin{minipage}{.8\textwidth}
\begin{it}
\noindent Instituto de Ciencias Nucleares \\
Universidad Nacional Aut\'onoma de M\'exico\\
Apdo. Postal 70-543, 04510 M\'exico, D.F., M\'EXICO \\[.5mm]
\end{it}
\texttt{chryss@nuclecu.unam.mx, eliokon@nuclecu.unam.mx 
\phantom{X}}
\end{minipage}
\\

\vspace*{3cm}
\noindent
\textsc{\large Abstract: }
We apply Lie algebra deformation theory to the problem of
identifying
the stable form of the quantum relativistic
kinematical algebra. As a warm up, given Galileo's conception of
spacetime as input, some modest computer code we wrote
zeroes in on the Poincar\'e-plus-Heisenberg algebra in 
about a minute. Further ahead, along the same path,
lies a
three dimensional deformation space, with an instability
double cone through its origin. We give physical as well as
geometrical arguments supporting our view that moment, rather
than position operators, should enter as generators in the Lie
algebra. With this identification, the deformation parameters
give rise to invariant length and mass scales. Moreover, standard
quantum relativistic kinematics of massive, spinless 
particles corresponds to non-commuting moment operators, a
purely quantum effect that bears no relation to spacetime
non-commutativity, in sharp contrast to earlier interpretations. 
\end{titlepage}
\setcounter{footnote}{1}
\renewcommand{\thefootnote}{\arabic{footnote}}
\setcounter{page}{2}
\noindent \rule{\textwidth}{.5mm}

\tableofcontents

\noindent \rule{\textwidth}{.5mm}
\section{Introduction}
\label{Intro}
A prevailing theme of the last decade or so in physics has
been the search for an algebraic signature of quantum
gravity. Lorentz symmetry violation, spacetime
non-commutativity and modified dispersion relations, among
other novelties, have been proposed as signals our antennas should be
tuned for, in the search for a scheme where quantum objects could 
be heavy too. 
Even before that, physicists exasperated by the darker side of
quantum field theory, sought their way out of the maze of
infinities in the form of a spacetime granularity that
would exorcise, the hope was, their ultraviolet nemeses. 

The usual suspect in many of these endeavors has been
the nature of spacetime, typically codified in a 
kinematical algebra. Accordingly, the above search has often
focused on the possible deformations of the
Lie algebra $\GPH$, \ie, the Poincar\'e algebra, extended by the
inclusion of the position operators and the Heisenberg
commutation relations, or one of its subalgebras. These attempts can be
roughly divided into three categories, based on the mathematical 
framework of their approach (or the absence thereof). The first
category comprises deformations of Lie type, where the
commutators of the generators are linear functions of the
same. There exists a well-developed mathematical formalism to
deal systematically with such deformations,
which has not always been used by physicists. We comment more 
extensively on this type of deformations below. In category number 
two enter quantum group type deformations, which are
generalizations of the classical group concept and form 
particular examples of Hopf algebras with a universal
$R$-matrix $\calR$, that solves the (universal) quantum 
Yang-Baxter equation%
\footnote{%
Not all Hopf-type deformations are known to posses 
a universal $R$-matrix, and some are known not to.%
}. 
The linearity of the Lie
case is lost%
\footnote{%
For a class of such algebras, an appropriate deformation of the 
concept of commutator
restores linearity (see, \eg.~\cite{Sch.Wat.Zum:93}).%
}, but the construction of the algebra is
canonical, given an $\calR$ with suitable properties. 
Extensive work in the eighties and nineties has provided a
solid mathematical background for these deformations, with
applications overflowing to an impressive list of fields. 
Finally, recent years have seen a plethora of articles loosely 
classified under the
generic misnomer (see~\cite{Ahl:03}) ``Doubly Special
Relativity'' (see, \eg,~\cite{Ame:02a}), which form the third 
category. A common feature among them, and in some sense the 
defining one, is that the commutators of the algebra are given by 
general analytic functions of the generators%
\footnote{%
The introduction of additional invariant scales cannot be considered 
as a defining characteristic of these deformations since, 
as is well-known, and as we are 
about to see, such scales are also introduced by the Lie-type 
deformations.%
}. To our knowledge, 
there is no well-defined mathematical framework guaranteeing
the self-consistency of these
deformations, not to mention their physical applicability, 
partly because they are not complete, \eg, the fate 
of the spacetime sector is often left unclear. Subsequent 
work~\cite{Kow.Now:02} showed
that endowing the above deformations with considerable
more (Hopf) structure, results in their identification with particular
forms of the $\kappa$-Poincar\'e Hopf algebra, proposed about a
decade ago~\cite{Luk.Now.Rue.Tol:91} (see also~\cite{Maj.Rue:94}).
Both of the last two categories suffer from serious physical 
problems
in the many-particle sector, \eg, in a consistent definition
of such basic quantities like the total momentum of a system of 
particles.

In this paper we deal with Lie-type deformations of standard
quantum relativistic kinematics. We undertook this project
with three main goals in mind:
\begin{enumerate}
\item
Emphasize the Lie algebra stability point of view and present in an 
accessible form the relevant mathematical apparatus, along the 
lines of Ref.~\cite{Vil:94}, which motivated the present work.
\item
Apply the formalism to the problem at hand 
to obtain a complete, detailed map of the deformation
territory in the vicinity of $\GPH$. 
\item
Interpret physically the generators of the algebra and
investigate the nature of the deformations.
\end{enumerate}
The structure of the paper was conceived accordingly, 
with each of the
subsequent three sections dealing with one of the above
goals. In
Sect.~\ref{LADatCoS} we give a self-contained review of the
standard Lie algebra deformation theory and explain why 
stable structures are more likely to prove useful in physical 
applications than unstable ones. The section ends with a
relatively detailed example, the passage from Galilean to
relativistic kinematics, illustrating the use of the
formalism, as well as the (alas, {\em a
posteriori}) predictive power of the stability point of view. 
Sect.~\ref{SQRK} contains a detailed analysis of the options
available in deforming $\GPH$. We take as our starting point
classical ($\hbar=0$) relativistic kinematics 
(an unstable algebra) and, 
with the help of some 
computer code we wrote, explore the various paths that lead to
stable algebras. We find that there is essentially one path,
its first stop introducing Heisenberg's relations. Thus, given
Galileo's conception of spacetime as input, our program
zeroes in on the Poincar\'e-plus-Heisenberg algebra $\GPH$ in 
about a minute. We find this motivating enough to inquire
about what lies further ahead.   
Following this path to its end, we find ourselves in a 
three-dimensional deformation space of stable Lie algebras, with a 
double instability cone through its origin. The section ends
with a description of relations between our work and earlier
treatments in the literature. 
This concludes the
mathematical part of the paper --- inferences about physics
will have to wait the physical identification of the
generators, which we undertake in Sect.~\ref{SPC}. There, we
argue that the position operators do not have the right
properties to serve as Lie algebra generators. 
In doing so, we are in disagreement 
with all previous works. Retracing
the steps that lead to the definition of the relativistic
center-of-momentum concept for a system of particles, we come to the
conclusion that the appropriate generators, in the case of a
massive, spinless particle, are the moment
operators, given essentially by the positions rescaled by the
mass operator for the particle. This shows that the algebra $\GQR$ 
of standard quantum relativistic kinematics, differs from
$\GPH$ and, in the above case, lies on the
instability cone. Furthermore, from the algebraic point of view, 
there is
a single deformation direction introducing non-commutativity
among the momenta. Sect.~\ref{CR} comments on the findings and 
outlines directions for future work. 
\section{Lie Algebra Deformations and the Concept of Stability}
\label{LADatCoS}
In this section we summarize the elements of standard Lie algebra
deformation theory that will be of use in the rest of the
paper. Our exposition follows mostly the  original source for this 
material~\cite{Nij.Ric:66,Nij.Ric:67}, as well as~\cite{Vil:94}.
Sect.~\ref{Coaecd} follows Ref.~\cite{Chr:01a}, echoing ideas
originating in the Batalin-Vilkovisky quantization (see,
\eg,~\cite{Sta:97}). Background information on Lie algebra
cohomology can be extracted (not without some effort) 
from~\cite{Che.Eil:48,Hoc.Ser:53}.
The foundations of deformation theory are laid out in
the classic~\cite{Ger:64}, while
plenty of newer material is contained in the
book-length~\cite{Ger.Sch:88}. An elegant generalization to 
bialgebra deformations was given in~\cite{Ger.Sch:90}, 
with still further generalizations to Drinfeld algebras, 
and much more, appearing in~\cite{Shn.Ste:93} --- this latter
reference also contains a rather comprehensive bibliography. 
An exposition of related material, with physical applications in mind,
can be found in~\cite{Azc.Izq:95}.
\subsection{Lie products}
\label{Lp}
We deal throughout with finite-dimensional real Lie algebras. These
are built on a (finite-dimensional) real vector space $V$, by
defining a bilinear antisymmetric {\em Lie product} map $\mu
\colon V \times V \rightarrow V$ that satisfies the Jacobi
identity,
\ble{Jacobiid}
\mu(x,\mu(y,z))=\mu(\mu(x,y),z)+\mu(y,\mu(x,z))
\, .
\ee 
This is usually written as a cyclic sum, a form that, in the
case at hand, obscures
its content. To clarify the latter, take as an example the
case where $x$ is a Lorentz group generator, $J_{\mu \nu}$, and
$y$, $z$ are other generators carrying Lorentz indices, say,
$Y_{\rho}$, $Z_{\sigma}$ respectively. Suppose 
$\mu(y,z)=\mu(Y_{\rho},Z_{\sigma})=W$. Substituting this in the \lhs{}
above, one finds that the Jacobi identity requires that the
transformation properties of $W$ under the Lorentz group are
derived solely from those of $Y_{\rho}$, $Z_{\sigma}$, \ie, in this
case, $W$ ought to transform as a second-rank covariant
tensor. Another way of saying this is that $\mu$ itself is a
Lorentz scalar, an observation that we use later on.

Given a basis $\{T_A\}$, $A=1,\ldots,n$ of $V$, the product
$\mu$ is specified by giving all vectors $\mu(T_A,T_B)$,
$1\leq A < B \leq n$. The coordinates of these vectors in the 
basis are, up to a factor of $i$, the {\em structure constants} of the algebra,
\ble{scdef}
[T_A,T_B] 
\equiv 
i \, \mu(T_A,T_B) 
=
i \, {f_{AB}}^C T_C
\, ,
\ee
which are antisymmetric in the lower two indices (a sum over
repeated indices is implied).
In the above equation we follow the standard physics practice of
expressing the (non-associative) Lie product as the commutator
$[\cdot,\cdot]$
\wrt{} an associative operator product, as well as the inclusion 
of an imaginary unit, related to the hermiticity of the 
generators. In terms of the
structure constants, the Jacobi identity becomes
\ble{antiJI}
  {f_{AR}}^S {f_{BC}}^R
+ {f_{BR}}^S {f_{CA}}^R
+ {f_{CR}}^S {f_{AB}}^R
=0
\, .
\ee
Relaxing for the moment this latter constraint, \ie, taking
into account only the antisymmetry in the lower two indices, one is
left with $N(n)=n^2(n-1)/2$ arbitrary constants ${f_{AB}}^C$, $A<B$. 
Consider now
the space $\mathbb{R}^N$, with each of the $f$'s ranging along
an axis.
For each value of $(A,B,C,S)$,~(\ref{antiJI}) describes a
quadratic hypersurface in this space. The intersection of
these hypersurfaces is
the space $\calL_n$ of all possible $n$-dimensional Lie
algebras
--- we sketch it as a surface in Fig.~\ref{allLA}. 
\setlength{\figurewidth}{.9\textwidth}
\begin{figure}
\begin{pspicture}(5mm,10mm)(\figurewidth,.7\figurewidth)
\setlength{\unitlength}{.25\figurewidth}
\psset{xunit=.25\figurewidth,yunit=.25\figurewidth,arrowsize=1.5pt
3}
\centerline{\raisebox{-.05\totalheight}{%
\psline[linewidth=.3mm]{->}%
(2.1,.9)(4.1,.6)
\psline[linewidth=.3mm]{->}%
(2.1,.9)(1.1,.5)
\includegraphics[angle=0,width=1.1\figurewidth]{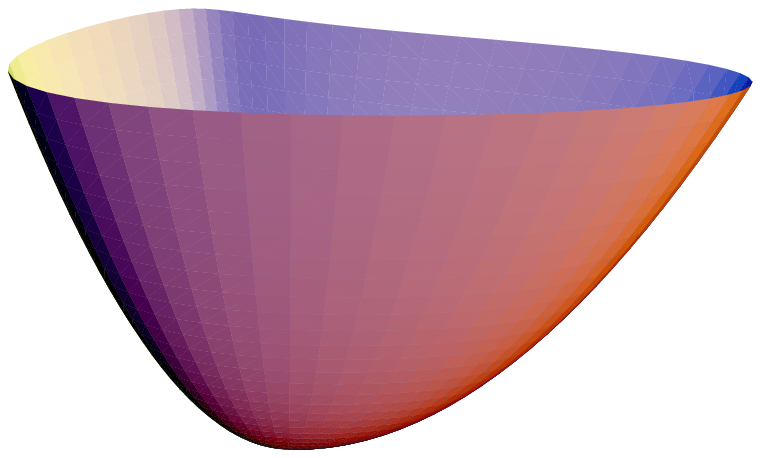}}}
\pscustom[fillstyle=solid,fillcolor=gray]{%
\psccurve(-2.6,1.67)(-2.4,1.54)(-2.12,1.4)(-2.2,1.3)(-2.46,1.33)}
\psdots[dotscale=1.3](-2.5,1.5)(-2.35,1.4)(-1.65,1.35)
\psline[linewidth=.3mm]{->}%
(-2.3,2.17)(-2.3,2.9)
\pscurve[linewidth=.3mm]{-}(-1.75,1)(-1.65,1.35)(-1.35,1.48)
\psline[linewidth=.3mm]{->}(-2.5,1.5)(-2.35,1.4)
\psline[linewidth=.3mm]{->}(-1.65,1.35)(-1.83,1.41)
\put(-2.8,1.85){\makebox[0cm][r]{{\Large $\calL_n$}}}
\put(-2.5,1.53){\makebox[0cm][r]{$P$}}
\put(-2.33,1.35){\makebox[0cm][l]{$P_M$}}
\put(-1.63,1.24){\makebox[0cm][l]{$Q$}}
\put(-1.80,1.43){\makebox[0cm][r]{$\psi_1$}}
\put(-3.22,.44){\makebox[0cm][r]{{\large ${f_{12}}^1$}}}
\put(-.1,.5){\makebox[0cm][r]{{\large ${f_{12}}^2$}}}
\put(-2.33,2.75){\makebox[0cm][r]{{\large ${f_{n-1,n}}^n$}}}
\rput(-2.5,1.16){\rnode{A}{$GL(n)$ orbit of $P$}}
\rput(-2.4,1.37){\rnode{B}{}}
\nccurve[angleA=70,angleB=-130,linewidth=.18mm]{->}{A}{B}
\rput(-1.55,1.85){\rnode{A}{$GL(n)$ orbit of $Q$}}
\rput(-1.36,1.49){\rnode{B}{}}
\nccurve[angleA=-45,angleB=90,linewidth=.18mm]{->}{A}{B}
\end{pspicture}
\capitem{\textsf{The space $\calL_n$ of $n$-dimensional Lie
algebras (sketch). $P$ is surrounded by equivalent points and hence,
$\calG_P \sim \calG_{P_M}$, for all $P_M$ 
sufficiently close to $P$. In contrast, in the
tangent space of $\calL_n$ at $Q$, there are directions that
lead
outside of the $GL(n,\mathbb{R})$ orbit $\text{Orb}(Q)$. 
$Q$ will move along these
directions when
$\psi_1$ in~(\ref{defcom}) is a non-trivial element of
$H^2(\calG_Q)$. Notice that, for all $n$, the surface passes 
through the
origin, which corresponds to the $n$-generator abelian algebra.}}
\label{allLA}
\end{figure}
Referring to
this figure, consider the point $P$ of $\calL_n$ --- it
corresponds
to the Lie algebra $\calG_P$, whose structure constants are
given by
the coordinates of $P$. Under a linear redefinition of the
generators via a $GL(n)$ matrix $M$,
\ble{linredef}
T'_A = {M_A}^B T_B
\, ,
\ee
the structure constants transform as
\ble{sctran}
{f'_{AB}}^C= {M_A}^R {M_B}^S {(M^{-1})_U}^C {f_{RS}}^U
\, ,
\ee
and $P$ moves to $P_M$. Clearly, no new physics is to be
expected
from such a redefinition, $\calG_P$ and $\calG_{P_M}$ being
isomorphic. What we are really interested in then,
from a physical point of view, is
not $\calL_n$ itself, but, rather, the space of equivalence classes
into which
$\calL_n$ splits under the above action of $GL(n)$, each class
being the $GL(n)$ orbit $\text{Orb}(P)$ of any point $P$ in the
class%
\footnote{\label{fn:isont}%
Notice that if one changes the structure constants from $f$ to
$f'$, as above, {\em without} changing the generators, one {\em does}
obtain new physics.%
}.
The crucial observation to be made here is that
there exist two
types of points in $\calL_n$: those that are completely
surrounded
by equivalent points (corresponding to isomorphic algebras) 
and those whose neighborhoods%
\footnote{%
$\calL_n$ inherits the natural topology of the structure
constants, \ie, that of the ambient $\mathbb{R}^N$.%
}
include non-equivalent points, sketched as $P$ and $Q$
respectively in
Fig.~\ref{allLA}. Any infinitesimal perturbation of the structure
constants of $\calG_P$ will
necessarily lead to an isomorphic Lie algebra --- the orbit
$\text{Orb}(P)$ is open in $\calL_n$. We call $\calG_P$ {\em
stable} or {\em rigid}. On the other hand, there
exist
infinitesimal perturbations of $\calG_Q$ that lead outside of
$\text{Orb}(Q)$ and, hence, to non-isomorphic algebras --- we
call $\calG_Q$ {\em (infinitesimally) unstable}.

In physical applications, structure constants are often given
by experimentally determined fundamental constants of the theory. 
The experimental errors involved render the position of the
corresponding algebra in $\calL_n$ uncertain. If the algebra 
employed is unstable, the physical predictions of the theory 
become ill defined, as they depend critically on the exact value 
of the structure constants, which is not known. 
Additionally, new measurements or improved data analysis, may move 
the algebra to a new position. If the algebra is stable, 
the physical theory based on it will maintain its qualitative 
validity. 
We conclude that stable algebras give rise to robust physics. 
\subsection{Deformations and $H^2$}
\label{DaH2}
Given a Lie algebra $\calG_0=(V,\mu_0)$, \ie, the Lie product 
of $X$, $Y \in V$ is supplied by
$\mu_0(X,Y) \equiv [X, \, Y]_0$. A {\em one-parameter
(formal) deformation}
of $\calG$ is given by the {\em deformed commutator}
\ble{defcom}
[X,\ Y]_t = [X, \, Y]_0
                + \sum_{m=1}^\infty \psi_m(X, \, Y)
\, t^m
\, ,
\ee
where $t$ is a formal parameter. The corresponding
$t$-dependent structure constants,
\ble{ft}
[T_A,\ T_B]_t= i \, {f_{AB}^t}^C T_C
\, ,
\ee
define a curve $P_t$ in
$\calL^n$, which passes through $P_0$ (corresponding to
$\calG_0$) at $t=0$. 
The \lhs{} of~(\ref{defcom})
is bilinear and antisymmetric, hence the $\psi_m$ on the
\rhs{} are
$\calG$-valued, bilinear antisymmetric maps
\ble{psim}
\psi_m: \quad V \times V \rightarrow V
\, ,
\qquad \qquad
\psi_m(X, \, Y) = - \psi_m(Y, \, X)
\, .
\ee
We will call such maps {\em 2-cochains} (over $V$), extending the 
definition in the natural way (\ie, via $p$-linearity and 
total antisymmetry) to  $p$-{\em cochains} $\psi^{(p)}$, 
which accept $p$ arguments%
\footnote{%
When the order $p$ of a cochain $\psi$ needs to emphasized, we will
write $\psi^{(p)}$.%
}. The vector space of $p$-cochains
over $V$
will be denoted by $C^p(V)$. Notice that the 1-cochains are
simply linear maps from $V$ to $V$, the
antisymmetry requirement being meaningless in this case. Also,
the space of 0-cochains is $V$ itself. 
Next, for an arbitrary Lie product $\mu$, we define a 
{\em
coboundary operator} $s_{\mu}$, which  maps 
$p$-cochains to $(p+1)$-cochains, 
$s_{\mu} \colon C^p \rightarrow C^{p+1}$,
according to
\begin{align}
\label{cobdef}
s_{\mu} \triangleright \psi^{(p)} (T_{A_0}, \ldots ,T_{A_{p}})
&=
\sum_{r=0}^{p} 
(-1)^{r} 
\mu \left(
T_{A_r}, \, 
\psi^{(p)} (T_{A_1}, \ldots ,\hat{T}_{A_r}, \ldots ,T_{A_{p}} ) 
\right)
\ff
 &\phantom{=}
{}+ \sum_{r<s} (-1)^{r+s} 
\psi^{(p)} \left(
\mu(T_{A_r}, \, T_{A_s}), 
\, T_{A_0}, \ldots ,\hat{T}_{A_r}, \ldots ,\hat{T}_{A_s},
\ldots , T_{A_{p}} 
\right)
\end{align}
(hats denote omitted terms).
For example, for $\phi \in C^1$ and $\psi \in C^2$, 
\bae
\label{saction}
s_{\mu} \triangleright \phi(A_1,A_2)
\fe
[A_1,\phi(A_2)]
-
[A_2,\phi(A_1)]
-
\phi([A_1,A_2])
\ff
s \triangleright \psi(A_1,A_2,A_3)
\fe
[A_1,\psi(A_2,A_3)]
-
[A_2,\psi(A_1,A_3)]
+
[A_3,\psi(A_1,A_2)]
\ff
 & &
{}-
\psi([A_1,A_2],A_3)
+
\psi([A_1,A_3],A_2)
-
\psi([A_2,A_3],A_1)
\, ,
\eae
where $\mu(X,Y)=[X,Y]$.
It can be shown that $s_{\mu}^2=0$, a result that relies on the 
Jacobi identity that
$\mu$ satisfies --- a compact proof is
given in Sect.~\ref{Tbwp}. 
The relevance of $s_{\mu}$ to the problem at hand becomes 
evident when one
imposes the Jacobi identity on the deformed commutator
in~(\ref{defcom}). Writing out this identity and evaluating
its $t$-derivative at $t=0$, one finds that
$\psi_1$ must satisfy $s_{\mu_0} \triangleright \psi_1=0$. 
We call a $p$-cochain $\psi^{(p)}$ annihilated by $s_{\mu}$, $s_{\mu}
\triangleright \psi^{(p)}=0$,  a
$p$-{\em cocycle}, and denote the vector space of $p$-cocycles by
$Z^p(V,s_{\mu})$.  What we have found above is that 
{\em infinitesimal deformations
of Lie algebras are generated by 2-cocycles}, the converse
being also true. 

It remains to determine which of the above infinitesimal 
deformations lead to 
isomorphic Lie algebras. As mentioned
already, isomorphic Lie algebras result 
from a linear
redefinition of the generators with some invertible matrix $M
\in GL(n,\mathbb{R})$ (see~(\ref{linredef})). 
In the case of a deformation, $M=M_t$
is $t$-dependent, with $M_0=I_n$ (the unit
$n \times n$ matrix). Then the deformed, $t$-dependent 
commutator, is given by
\ble{Mredef}
[X,Y]_t= M_t^{-1} [M_t X,M_t Y]_0
\, ,
\ee
for any $X$, $Y$ in $\calG_0$.
Taking $M_t$ in a neighborhood of the identity, 
$M_t=I_n + t Q$, with $t$ small,
one readily computes the corresponding first-order (in $t$) 
change to the commutator,
\bae
\label{comchange1}
[X, \, Y]_t
\fe
M_t^{-1} [M_t X, \, M_t Y]_0
\ff
 \fe
(I_n-tQ)[(I_n+tQ) X, \, (I_n+tQ) Y]_0
\ff
 \fe
[X, \, Y]_0 
+t \left(
[X, \, Q Y]_0 - [Y, \, Q X]_0 -Q[X, \, Y]_0
\right) +\calO(t^2)
\, .
\eae
But the linear map $Q \colon X=X^A T_A \mapsto Q X=X^R {Q_R}^S T_S$ is, 
as mentioned earlier, a 1-cochain.
Comparing the $\calO(t)$-term in the \rhs{}
of~(\ref{comchange1}) with the first of the examples
in~(\ref{saction}) shows that the $\calO(t)$-change in
the commutator, \ie, the 2-cochain $\psi_1$
in~(\ref{defcom}), is given by
\ble{psiQ}
\psi_1 = s_{\mu_0} \triangleright Q
\, .
\ee
We call a $p$-cochain $\psi^{(p)}$ that is in the image of
$s_{\mu}$,
$\psi^{(p)}=s_{\mu} \triangleright \phi^{(p-1)}$,
a {\em trivial $p$-cocycle}, or, a {\em $p$-coboundary}.
The vector space  of $p$-coboundaries will
be denoted by $B^p(V,s_{\mu})$. Since $s_{\mu}^2=0$, all 
coboundaries are
cocycles, $B^p \subseteq Z^p$. 
What the above result shows is
that {\em infinitesimal deformations of $\calG_0$ towards
isomorphic Lie algebras are generated by
2-coboundaries}. 

Conversely, assume that all 2-cocycles
are trivial. Given a deformed commutator as in~(\ref{defcom}),  
there exists a linear map (1-cochain) 
$\phi_1 \colon \calG_0 \rightarrow \calG_0$, such that 
the $\psi_1$ appearing in the \rhs{} of that equation is
given by $\psi_1= s_{\mu_0} \triangleright \phi_1$. 
Consider
now a linear redefinition of the generators by the matrix
$M_1=e^{-t \phi_1}$ and compute the new $t$-commutator
$[X,Y]'_t$.
The result is given by~(\ref{comchange1}), with the
substitutions $[X,Y]_t \rightarrow [X,Y]'_t$ and $[X,Y]_0
\rightarrow [X,Y]_t$,
\ble{comp}
[X,Y]'_t = [X,Y]_t -t 
\underbrace{s_{\mu_0} \triangleright \phi_1}_{\psi_1}(X,Y) +
\calO(t^2)
\, .
\ee
Using~(\ref{defcom}) to expand the \rhs{} in powers of $t$, 
we see that the term
linear in $t$ in $[X,Y]'_t$ cancels. Repeating the procedure one
may eliminate one by one all powers of $t$, thus bringing the
original $t$-commutator in coincidence with the undeformed one
$[X,Y]_0$, using nothing more than successive linear redefinitions 
of the
generators. We conclude that the two commutators define
isomorphic Lie algebras, the matrix giving the isomorphism
being $M=\ldots M_2 M_1$, with $M_m=e^{-t\phi_m}$ and
$s_{\mu_0}
\triangleright \phi_m = \psi_m$. 

We may summarize the contents of this section in the following
geometrical picture: the tangent space%
\footnote{%
We are assuming here that $P$ is not a singular point of
$\calL_n$ --- if that is not the case one may instead conclude
that
the Zariski tangent space to $\calL_n$ at $P$ is 
$Z^2$ (see~\cite{Nij.Ric:67}, \cite{Goz:88} p.{} 317).%
} $T_{P_0}\calL_n$ to
$\calL_n$ at $P_0$ is (isomorphic to) $Z^2$, the space of
2-cocycles. The subspace of $T_{P_0}\calL_n$ leading to
isomorphic Lie algebras, \ie, the tangent space to the
$GL(n)$-orbit $\text{Orb}(P)$ is $B^2$, the space of
2-coboundaries.  
To close the
familiar circle of definitions, we define the quotient space
$H^p \equiv Z^p/B^p$, in which two cocycles are identified if they 
differ
by a coboundary, as the $p${\em -th cohomology group}%
\footnote{%
$Z^p$, $B^p$, $H^p$ are all abelian groups with the group
composition given by addition.%
} of $\calG_0$. 
The non-trivial elements of
$H^2$ (if any) correspond to directions in $T_{P_0}\calL_n$ that lead
to Lie algebras infinitesimally close to
$\calG_0$ but non-isomorphic to it. 
A sufficient condition then for the stability of $\calG_0$ is
the
vanishing of its second cohomology group $H^2(\calG_0)$.
Whitehead's lemma states that this condition is satisfied by
all semisimple Lie algebras~\cite{Jac:79} --- we conclude that {\em
semisimple Lie algebras are stable}.
It is worth pointing out that the above is not a necessary
condition. As explained in Sect.~\ref{OaH3}, although a 
non-trivial 2-cocycle may exist,
obstructions originating in $H^3(\calG)$ can render it
non-integrable, in which case the corresponding {\em finite}
non-trivial deformation does not exist. Concrete examples of
stable Lie algebras with non-trivial $H^2$ have been
constructed, typically as semidirect products.
For example (see~\cite{Ric:67}), denote by $S$ the simple 
3-dimensional Lie algebra over 
$\mathbb{C}$ and by $\rho_n$ the irreducible representation of
weight $n$ of $S$ on $W\equiv \mathbb{C}^{2n+1}$. The
semidirect product $L_n=W \rtimes_{\rho_n} S$, for $n>5$ and odd, 
is a stable Lie
algebra, while its second cohomology group is non-trivial. To
deal with such cases, non-cohomological approaches have been
developed, relying on techniques of non-standard analysis. A
classification algorithm for stable Lie algebras
exists,
relying on a theorem that such algebras posses a standard
non-zero generator whose adjoint representation is
diagonalizable. Although tedious, the algorithm permits, in
principle, the classification of all stable Lie algebras, in
any dimension --- for more details we refer the reader
to~\cite{Anc:88,Goz:88}.
\subsection{The $\barwedge$ product} 
\label{Tbwp}
It turns out that calculations involving expressions
like~(\ref{Jacobiid}), or~(\ref{saction}), simplify
considerably when a particular product, the subject of this
section, is introduced among $p$-cochains~\cite{Nij.Ric:67}. 

Given a vector space $V$, put $\Alt^p(V) = C^{p+1}(V)$,
$p \geq -1$. Then for $\alpha \in \Alt^m(V)$, $\beta \in
\Alt^n(V)$, define the product $\alpha \barwedge
\beta \in \Alt^{m+n}(V)$ by
\ble{bwedgedef}
\alpha \barwedge \beta (X_0, \ldots, X_{m+n})
=
\sum_{\sigma}
\text{sgn}(\sigma) 
\, 
\alpha \big(
\beta(X_{\sigma(0)},\ldots,X_{\sigma(n)}),
X_{\sigma(n+1)}, \ldots, X_{\sigma(m+n)}
\big)
\, ,
\ee
where $\sigma$ ranges over all permutations such that
$\sigma(0) < \ldots < \sigma(n)$ and $\sigma(n+1) < \ldots <
\sigma (m+n)$ (these are known as {\em riffle shuffles with
cut at $n+1$}). When both $\alpha$ and $\beta$ are 2-cochains,
as will often be the case, the above formula reduces to
\ble{ab2cochain}
(\alpha \barwedge \beta)_{A B C}^{\phantom{A B C}T}=
\alpha_{RA}^{\phantom{RA}T} \beta_{BC}^{\phantom{BC}R} 
+\alpha_{RB}^{\phantom{RB}T} \beta_{CA}^{\phantom{CA}R} 
+\alpha_{RC}^{\phantom{RC}T} \beta_{AB}^{\phantom{AB}R} 
\, ,
\ee
where, for a $p$-cochain $\psi^{(p)}$,
\ble{psicomp}
\psi^{(p)}(T_{A_1}, \ldots, T_{A_p})
=
\psi_{A_1 \ldots A_p}^{\phantom{A_1 \ldots A_p}B}T_B
\, .
\ee
Notice that $\barwedge$ is non-associative, 
but satisfies instead
\ble{nonassoc}
(\gamma \barwedge \alpha) \barwedge \beta
-
\gamma \barwedge (\alpha \barwedge \beta)
=
(-1)^{mn}
\big(
(\gamma \barwedge \beta) \barwedge \alpha
-
\gamma \barwedge (\beta \barwedge \alpha)
\big)
\ee
(the {\em commutative-associative law}).
The {\em (graded) commutator} of $\alpha$, $\beta$
is defined as
\ble{gcommdef}
\llbracket \alpha,\beta \rrbracket
= \alpha \barwedge \beta - (-1)^{mn} \beta
\barwedge \alpha 
\, .
\ee
Consider now a Lie algebra $\calG=(V,\mu)$, 
$\mu \in C^2(V) =\Alt^1(V)$.
It is easy to see that the Jacobi identity for $\mu$,
Eq.~(\ref{Jacobiid}), can be put in the form 
\ble{Jacobimu2}
\mu \barwedge \mu 
= 
\frac{1}{2} \llbracket \mu,\mu \rrbracket
= 0
\ee
(the first
equality is an immediate consequence of~(\ref{gcommdef})).
Furthermore, the action of the coboundary operator $s_{\mu}$ on an
arbitrary $(p+1)$-cochain $\psi \in \Alt^p(V)$ is given by 
\ble{sactD}
s_{\mu} \triangleright \psi 
= 
(-1)^p \llbracket \mu, \psi \rrbracket 
\equiv 
(-1)^p D_{\mu} \psi
\, ,
\ee
\ie, $s_{\mu}$ is equal, up to a sign depending on the order of
the cochain it acts on, to the operator 
$D_\mu \equiv \llbracket \mu, \cdot \rrbracket$. 
Thus, all operations introduced
in earlier sections can be expressed in terms of the
$\barwedge$ product. 

It can be shown that the graded commutator of~(\ref{gcommdef})
satisfies a graded Jacobi identity,
\ble{gJacobi}
(-1)^{mp}  \llbracket \alpha, \, 
\llbracket \beta,  \, \gamma \rrbracket \rrbracket
+(-1)^{nm} \llbracket \beta,  \, 
\llbracket \gamma, \, \alpha \rrbracket \rrbracket 
+(-1)^{pn} \llbracket \gamma, \, 
\llbracket \alpha, \, \beta \rrbracket \rrbracket 
= 0
\, ,
\ee
where $\alpha \in \Alt^m(V)$, $\beta \in \Alt^n(V)$ and
$\gamma \in \Alt^p(V)$. This property, together with bilinearity
and graded antisymmetry, implies that $\Alt(V) \equiv
\bigoplus_n \Alt^n(V)$ is a graded Lie algebra. 

We derive now a number of interesting results, illustrating 
along the way the efficiency afforded by the formalism
introduced in this section.  
First, the proof of
$s_{\mu} \circ s_{\mu}=0$ may be given in a  simplified form. 
Up to an irrelevant sign, it
translates into $D_{\mu} \circ D_{\mu}=0$ and, for an arbitrary
$\alpha \in \Alt(V)$,
\ble{s2proof}
D_{\mu} \circ D_{\mu} \alpha 
= 
\llbracket \mu, \, 
\llbracket \mu, \, \alpha \rrbracket \rrbracket
=
\frac{1}{2} \llbracket \llbracket \mu, \,
\mu \rrbracket ,\, \alpha \rrbracket
=
0
\, ,
\ee
where the second equality follows from the graded Jacobi
identity for $\llbracket \cdot,\cdot \rrbracket$, 
Eq.~(\ref{gJacobi}), and the last one from the
Jacobi identity for $\mu$, Eq.~(\ref{Jacobimu2}).
Second, the equation for {\em finite} deformations may be
derived easily. If $\mu$ is a Lie product, $\mu'=\mu+\phi$
will also be one if 
$\llbracket \mu',\mu' \rrbracket=0$, from which one gets
immediately the {\em deformation equation}
\ble{defeq}
D_{\mu} \phi + \frac{1}{2} \llbracket \phi, \, \phi
\rrbracket=0
\, ,
\ee
which reduces to the cocycle condition for infinitesimal $\phi$. 
Third, Eq.~(\ref{gJacobi}) implies that $D_{\mu}$ is a graded
derivation in $\Alt(V)$, \ie,
\ble{Dmugder}
D_{\mu} \llbracket \alpha, \, \beta \rrbracket 
=
\llbracket D_{\mu} \alpha, \, \beta \rrbracket
+ (-1)^m \llbracket \alpha, \, D_{\mu} \beta \rrbracket
\, ,
\ee
where $\alpha \in \Alt^m(V)$ and $\beta \in \Alt(V)$. One may
then conclude that
if $\alpha$, $\beta$ are cocycles, $\alpha$,
$\beta \in Z(\Alt(V), D_{\mu})$, then so is
$\llbracket \alpha, \, \beta \rrbracket$, and that if,
additionaly, $\gamma$
is a coboundary, $\gamma \in B(\Alt(V), D_{\mu})$, 
then so is $\llbracket \alpha, \, \gamma
\rrbracket$. These two facts, in turn, imply that the quotient
space $H(\Alt(V),D_{\mu})$ is itself a graded Lie algebra.   
\subsection{Obstructions and $H^3$} 
\label{OaH3}
Given a Lie algebra $\calG=(V,\mu)$ and a deformation $\mu_t$,
\ble{mudef}
\mu_t=\mu+\phi_t \, ,
\qquad
\qquad
\phi_t = \sum_{n=1}^\infty \phi_n t^n
\, .
\ee
Then the deformation equation for $\phi_t$, Eq.~(\ref{defeq}),
implies an infinite sequence of equations for the $\phi_n$,
one for each power of $t$.
The equations corresponding to $t$, $t^2$ and $t^3$, are%
\footnote{%
Notice that all the $\phi_n$ are 2-cochains, so that
$\llbracket \phi_m,\phi_n \rrbracket =\llbracket
\phi_n,\phi_m \rrbracket=\phi_m \barwedge \phi_n+\phi_n
\barwedge \phi_m$.%
}
\begin{align}
\label{teq}
D_{\mu} \phi_1
&=
0
\\
\label{t2eq}
D_{\mu} \phi_2
&=
 -\frac{1}{2} \llbracket \phi_1, \, \phi_1 \rrbracket
\\
\label{t3eq}
D_{\mu} \phi_3
&=
-\llbracket \phi_1,\phi_2 \rrbracket
\, .
\end{align}
If $\phi_1$ is a 2-cocycle, as~(\ref{teq}) 
demands, then $\llbracket \phi_1, \phi_1 \rrbracket$ is a
3-cocycle, since $D_{\mu}$ is a (graded) derivation \wrt{} the 
$\llbracket \cdot,\cdot \rrbracket$ product. But then, 
(\ref{t2eq}) demands that this 3-cocycle be a
coboundary, which may not be the case if $H^3(V,D_{\mu})$ is
non-trivial. We see then that the existence of non-trivial
3-cocycles may render infinitesimal deformations ($\phi_1$
above) non-integrable. If $\llbracket \phi_1, \phi_1
\rrbracket$ is indeed a trivial 3-cocycle, so that~(\ref{t2eq}) 
admits a solution, an
obstruction may occur in the next step, \ie, in~(\ref{t3eq}), 
and so on. It can be shown that all of these
obstructions lie in $H^3$, so that, if $H^3$
is trivial, every non-trivial 2-cocycle is the first order
term of some finite deformation~\cite{Nij.Ric:67}.
 
The following remarks will prove useful:
\begin{enumerate}
\item
If a non-trivial 2-cocycle 
$\phi$ also satisfies 
$\llbracket \phi,\phi \rrbracket=0$, then it satisfies the
deformation equation~(\ref{defeq}). In that case, the
truncated deformation  $\mu_t=\mu +t \phi$ is a
Lie product for every $t$, if $\mu$ is one, regardless of the 
structure of $H^3$. 
\item
If there are several nontrivial 2-cocycles 
$\phi_i$ and all their anticommutators are zero, $\llbracket
\phi_i,\phi_j \rrbracket=0, \forall i,j$, then an arbitrary
linear combination of them also satisfies the finite
deformation equation, and the space of finite deformations
becomes a vector space, spanned by the $\phi_i$'s.  
\item
In both of the above cases, infinitesimal deformations along
non-trivial 2-cocycles are guaranteed to lead, as we saw earlier,
to non-isomorphic algebras. This is not necessarily
the case for finite deformations: the algebra $\mu +t\phi$
may become isomorphic, for particular finite values of $t$, to
the algebra $\mu$. Notice also that, in general, the algebras
$\mu +t\phi$,
for various finite values of $t$, may not be isomorphic among
themselves. The infinitesimal version of this is that the
algebras $\mu +t\phi$, for various (infinitesimal) values of 
$t$, are all
isomorphic, as long as $t$ does not change sign. $\mu +t\phi$
might well be non-isomorphic to $\mu -t\phi$, even for $t$
infinitesimal. 
\end{enumerate}
Interestingly enough, these scenarios are realized in the
stability analysis of the Galilean algebra, in
Sect.~\ref{GtoE}, as well as in that of the PH algebra, in 
Sect.~\ref{SQRK}.
\subsection{Coboundary operator as exterior covariant
derivative} 
\label{Coaecd}
It is obvious from the definition given above, that a
$p$-cochain
can be realized as a $\calG$-valued left invariant (LI)
$p$-form on the group manifold $G$ corresponding to $\calG$,
with the
generators $T_A$ now extended to LI vector fields. Denoting by
$\{\Pi^A\}$  the LI 1-forms on $G$ dual to the generators $\{T_B\}$,
\ble{PiTnot}
\ip{\Pi^A}{T_B}=\delta_B^{\phantom{B}A}
\, ,
\qquad 
(\text{with}
\quad
\ip{\Pi^{\mu \nu}}{T_{\rho \sigma}}
=
\delta_{\rho \sigma}^{\phantom{\rho \sigma}\mu \nu}
\equiv
 g_\rho^{\phantom{\rho}\mu} g_\sigma^{\phantom{\sigma}\nu} 
-g_\rho^{\phantom{\rho}\nu} g_\sigma^{\phantom{\sigma}\mu} 
)
\, ,
\ee
we write 
$\psi^{(p)}$ as
\ble{psiLI}
\psi^{(p)} 
\equiv 
\psi^B \otimes T_B
= 
\frac{1}{p!}
{\psi_{A_1 \ldots A_p}}^B \,
\Pi^{A_1} \ldots \Pi^{A_p}
\otimes T_B
\, .
\ee
Then the action of $s$ given in~(\ref{cobdef}) coincides with
that of an exterior covariant derivative $\nabla$,
\ble{nabladef}
\nabla (\psi^A \otimes T_A) 
= 
(d \psi^A + {\Omega^A}_B \psi^B) \otimes T_A
\, ,
\ee
with the
connection 1-form $\Omega$ given by
\ble{conndef}
{\Omega^A}_B = {f_{RB}}^A \Pi^R
\, ,
\qquad
(\mbox{\ie,}
\quad
\nabla_{T_A} T_B = [T_A, \, T_B])
\, .
\ee
The nilpotency of $s$ follows now from the vanishing of the
curvature 2-form
$\Theta = d\Omega + \Omega^2$, due to the Jacobi identity,
while
2-cocycles are covariantly constant $\calG$-valued LI
2-forms (see, \eg,~\cite{Chr:01a}). Notice that the requirement 
that $s \triangleright \psi^{(2)} = 0$, with
$\psi^{(2)}$ as in~(\ref{psiLI}), reduces to
\ble{JIlin}
  {f_{AR}}^S {\psi_{BC}}^R
+ {f_{BR}}^S {\psi_{CA}}^R
+ {f_{CR}}^S {\psi_{AB}}^R
+ {\psi_{AR}}^S {f_{BC}}^R
+ {\psi_{BR}}^S {f_{CA}}^R
+ {\psi_{CR}}^S {f_{AB}}^R
= 0
\, ,
\ee
which is, as expected, the first-order term, in $t$, of the Jacobi 
identity for the structure constants $f+t \psi$.
The use of the differential forms language permits 
writing out
cochains as geometrical objects, as in~(\ref{psiLI}), rather
than listing their components, a practice we adher to in the
following. 

The point of view sketched here has been further developed, in
the case of compact Lie algebras, in~\cite{Hol:90}, the
motivation there being the study of BRST cohomology. The
appropriate coboundary operator%
\footnote{%
Because of the compactness of the algebras studied
in~\cite{Hol:90}, 
the connection term in $\nabla$ is dropped --- otherwise the 
cohomology is trivial, as asserted by Whitehead's lemma.%
}, 
called there the BRST operator, 
is realized in terms of fermionic
coordinates and their dual derivatives. An involution of
the algebra of the latter, made possible by the invertibility of the
Killing form, gives rise to a dual object, the anti-BRST
operator, and a grade-preserving Laplacian. Further
generalizations, involving higher order invariant tensors of
the algebra,
have been explored in~\cite{Chr.Azc.Mac.Bue:99}. We have
developed similar techniques to deal with the non-compact
case, reinstating the connection term, and used them in one of 
our programming approaches ---
we defer further details to a future publication.   
\subsection{An example: the shortest path from Galileo to
Einstein}
\label{GtoE}
Consider the Galilean algebra $\GG$ of non-relativistic kinematics,
\be
[J_a,J_b]=i \, \eabc  J_c
\, ,
\qquad
\qquad
[J_a,K_b]=i \, \eabc  K_c
\, ,
\qquad
\qquad
[K_a,K_b]=0
\, ,
\ee
where $J_a$, $K_a$, $a=1,2,3$, are the generators of rotations and 
boosts,
respectively, and indices are raised and lowered with the unit
metric. The 2-cochain $\mu$ that corresponds to this Lie
product is, in the language of the preceding section,
\ble{2cochainG}
\mu = 
\frac{1}{2} 
\eabc
\Pi^a \Pi^b \otimes J_c
+ 
\eabc
\Pi^a \Pi^{\bar{b}} \otimes K_c 
\, .
\ee
We adopt here the convention that, in 1-forms, unbarred indices 
refer to rotations, while barred ones to boosts, so that, \eg,
$\ip{\Pi^{\bar{a}}}{K_b}=\delta_b^{\phantom{b}a}$ (notice that
bars are important in forms but make no difference in
Kronecker deltas or in the summation convention).
By an argument based on the observation made after
Eq.~(\ref{Jacobiid}), we conclude that only scalar (under
rotations) cochains need be considered.  
We simplify further the notation taking advantage of the fact
that, due to the limited number of generators and invariant
tensors, a simple listing of the nature of the 1-forms and
generators that enter in any given cochain, of up to second
degree, is sufficient to
reconstruct it (there is only one way to contract the
indices). For example, $\mu$ above is given by
\ble{muGdef}
\mu=\chiJJJ+\chiJKK
\, ,
\qquad
\qquad
\text{where }
\quad
\chiJJJ \equiv
\frac{1}{2} \eabc \Pi^a \Pi^b \otimes J_c
\, ,
\quad
\chiJKK
\equiv \eabc \Pi^a \Pi^{\bar{b}} \otimes K_c
\ee
(a factor of
$1/p!$ is included whenever $p$ 1-forms of the same
type are multiplied).    

We inquire now about the stability of this algebra. 
The most general scalar 1-cochain is given by
\be
\phi 
= 
  \alpha_1 \, \phiJJ
+ \alpha_2 \, \phiKJ
+ \alpha_3 \, \phiJK
+ \alpha_4 \, \phiKK
\, ,
\ee
with $\phiJJ=\Pi^a \otimes J_a$ \etc.
Applying $\nabla$ to obtain the most general 2-coboundary we
get
\be
\nabla \phi = 
\alpha_1 \left( \chiJJJ + \chiJKK \right)
+ 2 \alpha_2 \, \chiKKK 
+   \alpha_3 \, \chiJJK 
\, .
\ee
On the other hand, the most general scalar 2-cochain is given
by
\be
\chi = 
  \beta_1 \, \chiJJJ
+ \beta_2 \, \chiJJK
+ \beta_3 \, \chiJKJ
+ \beta_4 \, \chiJKK
+ \beta_5 \, \chiKKJ
+ \beta_6 \, \chiKKK 
\, .
\ee
We set $\nabla \chi =0$ to obtain
\be
\nabla \chi 
= 
\left( \beta_1 - \beta_4 \right) \Psi_1 
+ \beta_3 \, \Psi_2 
= 0
\ee
with
\ble{psi12}
\Psi_1 
=
\Pi^a \Pi^b \Pi_{\bar{b}} \otimes K_{\bar{a}} 
\, ,
\qquad
\Psi_2 
=
\Pi^a \Pi^b \Pi_{\bar{a}} \otimes J_b
+ \Pi^{\bar{b}} \Pi^a \Pi_{\bar{a}} \otimes K_b 
\, .
\ee
We conclude that $\beta_1=\beta_4$ and $\beta_3=0$, so that 
the most general 2-cocycle is given by
\be
\tilde{\chi} 
= 
c_1 \left(
\chiJJJ + \chiJKK 
\right)
+ c_2 \, 
\chiJJK 
+ c_3 \,
\chiKKJ 
+ c_4 \, 
\chiKKK 
\, ,
\ee
with arbitrary $c_i$.
Comparison of $\tilde{\chi}$ with $\nabla \phi$ shows that 
only $\chiKKJ$ is a non-trivial 2-cocycle, giving for the
second cohomology group
\be
H^2(\GG)=\{ [0], [\chiKKJ] \} 
\, .
\ee
Accordingly, $\GG$ is infinitesimally unstable.
By noting that 
$\llbracket \chiKKJ, \chiKKJ \rrbracket =0$, we
conclude that
$\mu_t=\mu +t \chiKKJ$ yields a one-parameter deformation of the
algebra for finite $t$ (see the comment at the end of 
Sect.~\ref{OaH3}). 
A look at the form of $\chiKKJ$ shows that the
deformation only adds a rotation generator in the \rhs{} of
the $K$-$K$ commutator,
\ble{KKdef}
[K_a,K_b]_t=i \, t \eabc J_c
\, ,
\ee
leaving the rest of the commutators intact. The Lorentz
algebra,
describing relativistic kinematics, sits at 
$t=-\frac{1}{c^2}$, where $c$ is the velocity of light and,
being semisimple, it is stable. 
\section{Stable Quantum Relativistic Kinematics}
\label{SQRK}
Hopefully, the above example will have aroused the interest of
the reader enough to follow us as we embark on the search for
a stable Lie algebra, encompassing relativistic and quantum
effects.   
Our starting point is the fourteen generator
Poincar\'e-plus-positions algebra
\begin{align}
\label{LLcr}
[J_{\mu \nu}, \, J_{\rho \sigma}] 
&= 
i \, \big(
g_{\mu \sigma} J_{\nu \rho}
+ g_{\nu \rho} J_{\mu \sigma}
- g_{\mu \rho} J_{\nu \sigma}
- g_{\nu \sigma} J_{\mu \rho}
\big)
\\
\label{LPcr}
[J_{\rho \sigma}, \, P_\mu]
&=
i \, \big(
g_{\mu \sigma} P_\rho - g_{\mu \rho} P_\sigma
\big)
\\
\label{LZcr}
[J_{\rho \sigma}, \, Z_\mu]
&=
i \, \big(
g_{\mu \sigma} Z_\rho - g_{\mu \rho} Z_\sigma
\big)
\, ,
\end{align}
augmented by a central generator $M$, to appear later in the
\rhs{} of the Heisenberg commutator.  We follow the practice of
omiting all zero commutators,
the metric used is $g=\text{diag}(1,-1,-1,-1)$ and $c$, the speed
of light, is taken equal to 1.
The resulting fifteen
generator algebra, describing classical ($\hbar=0$)
relativistic kinematics (plus the extra generator $M$) we call
$\GCR$ (``Classical Relativity''). 
The reader might want to identify 
$J_{\mu \nu}$ with the
Lorentz algebra generators and $P_\mu$ with the momenta (and,
even, $Z_\mu$ with the positions) but we will focus initially 
on the
strictly algebraic problem of stability, and only digress on
interpretational aspects, which hold some surprises, 
in Sect.~\ref{SPC}.

The 2-cochain $\muCR$, corresponding to $\GCR$, is given by
\ble{muCRdef}
\muCR = \frac{1}{2}
\Pi^{\alpha \rho}\Pi_{\rho}^{\phantom{\rho} \beta} \otimes
J_{\alpha \beta} 
+ \Pi^{\alpha \rho}
\Pi_{\rho} \otimes P_{\alpha}
+ \Pi^{\alpha \rho}
\Pi_{\dot{\rho}} \otimes Z_{\alpha}
\, .
\ee
A straightforward calculation shows that $\llbracket \muCR,\,
\muCR \rrbracket =0$, confirming that the Jacobi identity is
satisfied in $\GCR$.
\subsection{Calculation of \mvb $H^2(\GCR)$ \mvn} 
\label{CoH2}
We computed the second cohomology group  $H^2(\GCR)$ with 
the help of {\em MATHEMATICA}. We did this in two independent ways. 
In the first one, the components of cochains were calculated
explicitly, one-by-one, while in the second a symbolic
approach was followed, dealing, \eg, with sums of the form
$\Pi^{\mu}\Pi^{\nu}\otimes J_{\mu\nu}$ without expanding them
further. The first approach has the advantage of generality,
as it can deal, practically without further fine-tuning, with
any Lie algebra --- details of the calculation are given in the
appendix (see Sect.~\ref{H2GCR}). The second approach is 
generally faster, at
the price of adjustments needed every time a new object (\eg,
an invariant tensor) is introduced. In both approaches, the
remark made after Eq.~(\ref{Jacobiid}) shows that we may
consider only Lorentz scalars, drastically reducing the
workload. We have, nevertheless, implemented this only in our
second approach, to keep the first as general as possible. The
result of both calculations is
\ble{H2GCRdef}
H^2(\GCR)=\{[0],[\psiH],[\psiPMZ],[\psiZMP],
[\psiPMP],[\psiZMZ] \}
\, ,
\ee
\ie, there are five nontrivial generators,
 with representatives given by
\begin{align}
\label{psiHdef}
\psiH
&=
\Pi^\mu \Pi_{\dot{\mu}} \otimes M
\\
\label{psiPMZdef}
\psiPMZ
&=
\Pi^\mu \Pi^M \otimes Z_{\mu}
\\
\label{psiZMPdef}
\psiZMP
&=
\Pi^{\dot{\mu}} \Pi^M \otimes P_{\mu}
\\
\label{psiPMPdef}
\psiPMP
&=
\Pi^{\mu} \Pi^M \otimes P_{\mu}
\\
\label{psiZMZdef}
\psiZMZ
&=
\Pi^{\dot{\mu}} \Pi^M \otimes Z_{\mu}
\, .
\end{align}
As in Ex.~\ref{GtoE}, we adopt a compact notation where undotted 
indices in forms refer
to $P$'s, dotted ones to $Z$'s, so that, \eg,
$\ip{\Pi^{\dot{\mu}}}{Z_\nu}=\delta_{\nu}^{\phantom{\nu}\mu}$.
As before, dots make no difference in Kronecker delta's or
epsilon tensors. With a slight abuse of notation, $\Pi^M$
denotes the 1-form that detects the generator $M$.
\subsection{Finite deformations of \mvb $\GCR$ \mvn} 
\label{FdGCR}
Each of the cocycles in Eqs.~(\ref{psiHdef})--(\ref{psiZMZdef}) 
represents a direction of a possible infinitesimal deformation. 
For example, the first of these, $\psiH$, when added
to $\muCR$, adds the Heisenberg commutator to $\GCR$, while
each of the other four renders $M$ noncentral. There are two  
questions that
arise now:
\begin{enumerate}
\item 
Are these infinitesimal deformations
integrable?
\item
Are linear combinations of these infinitesimal deformations
integrable?
\end{enumerate}
To this end, we compute the commutators among the
$\psi$'s and find that the only nonzero ones are those between
$\psiH$ and the rest of the $\psi$'s --- it will
prove convenient in what follows to use the linear
combinations 
$\psi_- = \psiZMZ -\psiPMP$ 
and 
$\psi_+=\psiZMZ +\psiPMP$,
\begin{align}
\label{HpsiPMZ}
\llbracket \psiH,\, \psiPMZ \rrbracket
&=
-\Pi^{\mu} \Pi^{\nu} \Pi_{\dot{\nu}} \otimes Z_{\mu}
\\
\label{HpsiZMP}
\llbracket \psiH,\, \psiZMP \rrbracket
&=
\Pi^{\nu} \Pi^{\dot{\mu}} \Pi_{\dot{\nu}} \otimes P_{\mu}
\\
\label{Hpsim}
\llbracket \psiH,\, \psi_- \rrbracket
&=
\Pi^{\mu} \Pi^{\nu} \Pi_{\dot{\nu}} \otimes P_{\mu}
+\Pi^{\nu} \Pi^{\dot{\mu}} \Pi_{\dot{\nu}} \otimes Z_{\mu}
\\
\label{Hpsip}
\llbracket \psiH,\, \psi_+ \rrbracket
&=
\Pi^{\nu} \Pi^{\dot{\mu}} \Pi_{\dot{\nu}} \otimes Z_{\mu}
-\Pi^{\mu} \Pi^{\nu} \Pi_{\dot{\nu}} \otimes P_{\mu}
-2 \Pi^M \Pi^{\nu} \Pi_{\dot{\nu}} \otimes M
\, .
\end{align}
Regarding the first question above, the fact that the diagonal
commutators are all zero implies that $\muCR +t \psi_A$, for
$t$ finite, gives a deformation of $\GCR$, where $\psi_A$ is
any of the five generators of $H^2(\GCR)$ given above,
Eqs.~(\ref{psiHdef})--(\ref{psiZMZdef}). For the
second question, the fact that the commutators among the last
four $\psi$'s are all zero implies that any linear combination
of these $\psi$'s gives rise to a finite deformation as above.
The case of deformations that mix $\psiH$ with the other four
generators needs special treatment. We consider an
infinitesimal deformation along the 2-cocycle $\phi_1$,
\ble{phi1def}
\phi_1= q \psiH +\beta_1 \psiPMZ + \beta_2 \psiZMP + \beta_-
\psi_- +\beta_+ \psi_+
\, .
\ee
One easily checks that $\llbracket \phi_1,\phi_1 \rrbracket
\neq 0$ (the relevant anticommutators are given in
Eqs.~(\ref{HpsiPMZ})--(\ref{Hpsip}) above), so that
(\ref{t2eq}) is not trivially satisfied. For the above
mentioned anticommutators we find that the first three are
trivial,
\begin{align}
\label{HpsiPMZtr}
\llbracket \psiH,\, \psiPMZ \rrbracket
&=
-D_{\muCR} \psiPPJ
\\
\label{HpsiZMPtr}
\llbracket \psiH,\, \psiZMP \rrbracket
&=
D_{\muCR} \psiZZJ
\\
\label{Hpsimtr}
\llbracket \psiH,\, \psi_- \rrbracket
&=
-D_{\muCR} \psiPZJ
\, , 
\end{align}
where
\begin{align}
\label{psiPPLdef}
\psiPPJ
&=
 \frac{1}{2} \Pi^{\mu} \Pi^{\nu} \otimes J_{\mu \nu}
\\
\label{psiZZLdef}
\psiZZJ
&=
\frac{1}{2} \Pi^{\dot{\mu}} \Pi^{\dot{\nu}} \otimes J_{\mu \nu}
\\
\label{psiPZLdef}
\psiPZJ
&=
\Pi^{\mu} \Pi^{\dot{\nu}} \otimes J_{\mu \nu}
\, . 
\end{align}
On the other hand, 
$\llbracket \psiH,\, \psi_+ \rrbracket$ turns out to be
non-trivial. Accordingly, the infinitesimal
deformation generated by $\phi_1$ is integrable if, and
only if, $\beta_+=0$. In that case, Eq.~(\ref{t2eq}) is
satisfied with
\ble{phi2def}
\phi_2 = q \beta_1 \psiPPJ -q \beta_2 \psiZZJ 
+q \beta_- \psiPZJ
\, .
\ee
With an eye on~(\ref{t3eq}), we compute $\llbracket
\phi_1,\phi_2 \rrbracket$ and find that it vanishes, so that
$\phi_3=0$ (see (\ref{t3eq})). Also, $\llbracket \phi_2,\phi_2
\rrbracket=0$, implying that $\phi_4$, and all higher order
$\phi_n$'s vanish, and
the finite deformation truncates at second order,
\begin{align}
\muCR +\phi_t
&=
\muCR +\phi_1 t+\phi_2 t^2
\nonumber
\\
 &=
\muCR + (
q \psiH +\beta_1 \psiPMZ + \beta_2 \psiZMP +\beta_- \psi_-)t
+q(\beta_1 \psiPPJ - \beta_2 \psiZZJ + \beta_- \psiPZJ)t^2
\, .
\label{findefCR}
\end{align}
Without loss of generality, we may put $t=1$ and write the
result as
\ble{findefCR2}
\muCR +\phi_{t=1}=
\muCR + 
q \psiH 
+ \beta_1 (\psiPMZ + q \psiPPJ)
+ \beta_2 (\psiZMP - q \psiZZJ)
+ \beta_- (\psi_- +  q \psiPZJ)
\, ,
\ee
a form that will prove useful shortly.
\subsection{Heisenberg's route: 
the algebra \mvb $\GPH(q)$ \mvn} 
\label{Qe}
We want to explore here what happens if one follows, along
with Heisenberg, the
historical route and only deforms $\GCR$ along $\psiH$. 
We consider, accordingly, the stability of the algebra 
$\GPH(q)$ (``Poincar\'e plus Heisenberg''), with corresponding
2-cochain $\muPH(q)$ given by
\ble{muPHdef}
\muPH(q) = \muCR + q \psiH
\ee
(we assume henceforth that $q \neq 0$). 
The commutators defining it are given by
Eqs.~(\ref{LLcr})--(\ref{LZcr}), plus the  
Heisenberg commutator --- for the sake of locality we collect 
them all here,
\begin{align}
\tag{\ref{LLcr}$'$}
[J_{\mu \nu}, \, J_{\rho \sigma}] 
&= 
i \, \big(
g_{\mu \sigma} J_{\nu \rho}
+ g_{\nu \rho} J_{\mu \sigma}
- g_{\mu \rho} J_{\nu \sigma}
- g_{\nu \sigma} J_{\mu \rho}
\big)
\\
\tag{\ref{LPcr}$'$}
[J_{\rho \sigma}, \, P_\mu]
&=
i \, \big(
g_{\mu \sigma} P_\rho - g_{\mu \rho} P_\sigma
\big)
\\
\tag{\ref{LZcr}$'$}
[J_{\rho \sigma}, \, Z_\mu]
&=
i \, \big(
g_{\mu \sigma} Z_\rho - g_{\mu \rho} Z_\sigma
\big)
\\
\label{HeisenLie}
[P_{\mu},Z_{\nu}]
&=
i \, q \, g_{\mu\nu} M
\, .
\end{align}
We will have more to say about~(\ref{HeisenLie}) in Sect.~\ref{SPC} 
--- 
for the moment, we ask the reader to accept it as a reasonable
(\ie, covariant and of Lie-type) form of the familiar
Heisenberg relation.
As in the previous case, of
$\GCR$, we first tackle the purely algebraic
problem of stability, and leave questions of physical
interpretation for Sect.~\ref{SPC}. Meanwhile, 
the temptation should be resisted to consider $\GPH$ as the algebra of 
``quantum relativistic kinematics'' --- it is argued later on that 
it is not. 

We find that $H^2(\GPH(q))$ is nontrivial,
\ble{H2GPHdef}
H^2(\GPH(q))=\{
[0],[\zeta_1],[\zeta_2],[\zeta_3]
\}
\, ,
\ee
where
\begin{align}
\label{phiPPLdef}
\zeta_1
&=
\psiPMZ + q \psiPPJ
\\
\label{phiZZLdef}
\zeta_2
&=
- \psiZMP + q \psiZZJ
\\
\label{phiPZLdef}
\zeta_3
&=
\psi_- +q \psiPZJ
\, .
\end{align}
$\psiH$ itself is still a cocycle, albeit a trivial
one now (for all nonzero $q$). This means that, starting at $\GPH(q)$, 
and moving along 
$\psiH$, one arrives at isomorphic algebras. But
moving along $\psiH$ amounts to changing the value
of $q$, without changing its sign. 
We conclude that the algebras $\GPH(q)$, for all
nonzero values of $q$, of the same sign, are isomorphic. On
the other hand,
one can easily change the sign of $q$ by a redefinition of the
generators, \eg, by rescaling all $Z$'s by some negative
number, or exchanging $P$ and $Z$ (notice that in both examples, the
corresponding matrix that effects the redefinition has
positive determinant, \ie, it lies in the connected component of
$GL(15,\mathbb{R})$).  
The upshot of all this is that all $\GPH(q)$, for nonzero $q$, are
isomorphic. 
\subsection{Finite deformations of \mvb $\GPH(q)$ \mvn} 
\label{FdGPH}
Reasoning as in Sect.~\ref{FdGCR}, we compute the commutators
among the $\zeta$'s, and find that they all vanish.
Accordingly, every linear combination 
of the above cocycles, added to $\muPH$, provides a finite
deformation of $\GPH$. For a generic combination 
$\zeta(\vec{\alpha})$,
\ble{phigendef}
\zeta(\vec{\alpha})
=
 \alpha_1 \zeta_1
+\alpha_2 \zeta_2
+\alpha_3 \zeta_3
\, ,
\ee
the 2-cochain
$\mu(q,\vec{\alpha})=\muPH(q)+\zeta(\vec{\alpha})$, is clearly 
identical to the one found before, Eq.~(\ref{findefCR2}), with
the identifications $(\beta_1,\beta_2,\beta_-) \mapsto
(\alpha_1,-\alpha_2,\alpha_3)$. We have arrived then at the same 
result, either deforming $\GCR$ along a direction that truncates 
to second order in the deformation parameter, or by performing 
two succesive deformations (with intermediate stop at $\GPH(q)$),
each truncating to first order. The corresponding deformed algebra 
is given by the
commutators of $\GCR$, Eqs.~(\ref{LLcr})--(\ref{LZcr}) 
(notice that the Heisenberg commutator
is not included), plus the following
\begin{align}
\label{PZcomm}
[P_\mu, Z_\nu]
&=
 i \, q g_{\mu \nu} M
+i \, q \alpha_3 J_{\mu \nu}
\\
\label{PPcomm}
[P_\mu, P_\nu]
&=
i \, q \alpha_1 J_{\mu \nu}
\\
\label{ZZcomm}
[Z_\mu, Z_\nu]
&=
i \, q \alpha_2 J_{\mu \nu}
\\  
\label{PMcomm}
[P_\mu,M]
&=
-i \, \alpha_3 P_\mu 
+ i \, \alpha_1 Z_\mu
\\
\label{ZMcomm}
[Z_\mu,M]
&=
-i \, \alpha_2 P_\mu
+ i \, \alpha_3 Z_\mu
\, .
\end{align}
We denote the resulting algebra by $\GPH(q, \vec{\alpha})$. 
We see that, for a generic deformation, the $P$'s cease to
commute among themselves, and so do the $Z$'s, $M$ is no
longer central, while the Heisenberg commutator receives an
additional term, proportional to $J_{\mu \nu}$. 
\subsection{The instability cone} 
\label{Tic}
Relevant questions that emerge now
are: 
\begin{enumerate}
\item
Are the above deformations $\GPH(q,\vec{\alpha})$, for various 
values of $\vec{\alpha}$, (finally) stable?
\item
Are there deformations that are isomorphic among themselves?
\end{enumerate}
To answer the first question, we compute, as always, the
second cohomology group and find 
\ble{H2GPHqa}
H^2\big( \GPH(q,\vec{\alpha}) \big) = 
\begin{cases}
\{[0] \}& \text{if $\alpha_3^2 \neq \alpha_1 \alpha_2$} \\
\{[0], [\chi] \}& \text{if $\alpha_3^2 = \alpha_1 \alpha_2$}
\end{cases}
\, ,
\ee
where $\chi = \zeta_1 +\zeta_2$ satisfies 
$\llbracket \chi,\chi \rrbracket=0$.
$\GPH(q,\vec{\alpha})$ is, accordingly, stable
everywhere outside the instability surface $\alpha_3^2 = \alpha_1
\alpha_2$ in $\alpha$-space. The latter represents a double cone 
with the apex
at the origin and its axis along the first diagonal in the
$\alpha_1$-$\alpha_2$ plane, parallel to $\chi$ 
(see Fig.~\ref{instPlot}). We will
refer to the various regions of $\alpha$-space with their
relativistic nicknames (``future'', ``past'', \etc), with the
future including the positive $\alpha_1$-$\alpha_2$ quadrant.
Notice that, off the cone, $\chi$ is 
a trivial cocycle, $\chi = \nabla \xi$, and $\xi$
has a pole on the cone.   
Regarding the second question above, 
from the fact that each algebra outside the light cone is
isomorphic to all algebras in its neighborhood, we conclude
that all algebras in, say, the future, are isomorphic among
themselves (similarly for the past and the elsewhere). A
slight refinement of the argument, using the fact that tangent
vectors to the cone are trivial cocycles, leads to the
conclusion that all algebras in, say, the future cone, are
isomorphic among themselves (similarly for the past cone),
with the apex, \ie, $\GPH(q,\vec{0}) \equiv \GPH(q)$, 
in a class by itself. 
In conclusion, {\em there are
six equivalence classes of algebras, given by the various
regions the $\alpha$-space is divided into by the double light cone}. 
\setlength{\figurewidth}{\textwidth}
\begin{figure}
\begin{pspicture}(-.1\figurewidth,.01\figurewidth)%
(.9\figurewidth,.7\figurewidth)
\setlength{\unitlength}{.25\figurewidth}
\psset{xunit=.25\figurewidth,yunit=.25\figurewidth,arrowsize=1.5pt
3}
\centerline{\raisebox{-.22\totalheight}{%
\includegraphics[angle=0,width=1.3\figurewidth]{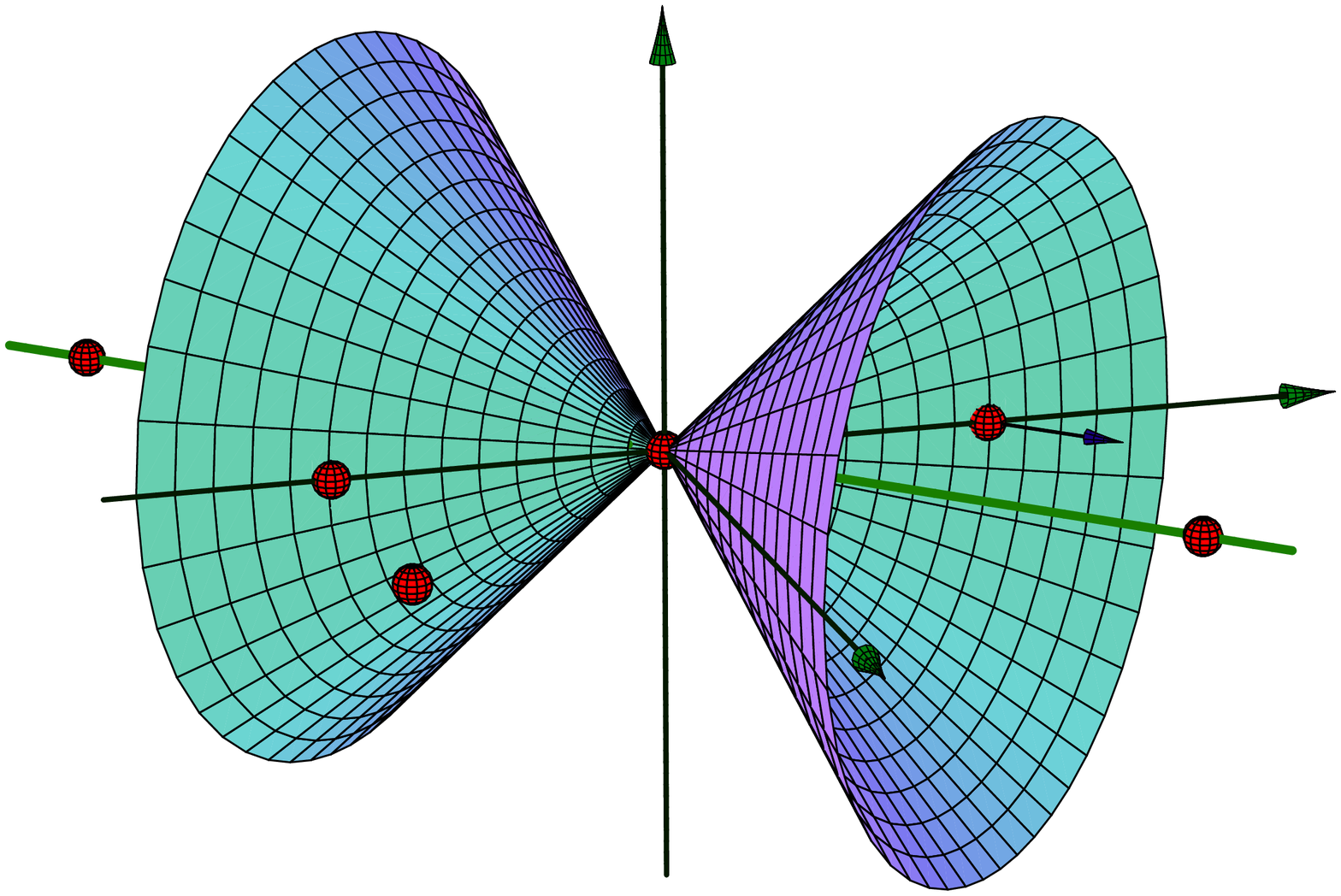}}}
\rput(-.75,2){\rnode{A}{{\Large $\calG_{\text{QR}}$}}}
\rput(-1.45,1.6){\rnode{B}{}}
\nccurve[angleA=190,angleB=30,linewidth=.25mm]{->}{A}{B}
\put(-.55,1.32){\makebox[0cm][r]{{\Large $\mathfrak{so}(1,5)$}}}
\rput(-4.24,.85){\rnode{E}{{\Large $\mathfrak{so}(2,4)$}}}
\rput(-3.3,1.06){\rnode{F}{}}
\nccurve[angleA=15,angleB=190,linewidth=.25mm]{->}{E}{F}
\rput(-2.2,2.1){\rnode{I}{{\Large $\GPH(q)$}}}
\rput(-2.42,1.63){\rnode{J}{}}
\nccurve[angleA=240,angleB=70,linewidth=.25mm]{->}{I}{J}
\put(-4.07,1.88){\makebox[0cm][r]{{\Large $\mathfrak{so}(3,3)$}}}
\put(-.54,1.74){\makebox[0cm][l]{{\Large $\alpha_2$}}}
\put(-2.41,2.73){\makebox[0cm][l]{{\Large $\alpha_3$}}}
\put(-1.08,1.48){\makebox[0cm][l]{{\Large $\chi$}}}
\put(-.45,.95){\makebox[0cm][r]{{\Large \textsc{Future}}}}
\put(-4.1,2.2){\makebox[0cm][r]{{\Large \textsc{Past}}}}
\put(-2.55,.5){\makebox[0cm][r]{{\Large \textsc{Elsewhere}}}}
\end{pspicture}
\capitem{\textsf{%
The $(\alpha_1,\alpha_2,\alpha_3)$ deformation space of
$\GPH(q)$, with a representative of each of the six
equivalence classes drawn (these are, in relativistic
parlance, the future and past 
light-cones, the apex, the future, the past and the elsewhere). 
The two cones and the apex at the
origin correspond to unstable algebras -- the rest of the
space to stable ones. 
For all classes, a representative exists with $\alpha_3=0$ 
(the little spheres denote such representatives).
With the identification of $Z_{\mu}$ with the
moment operator (see Sect.~\ref{SPC}), the origin corresponds to
$\GPH(q)$, while $\GQR$, the Lie algebra of standard quantum
relativistic kinematics, lies on the
future cone, at $\vec{\alpha}=(0,q,0)$.
Stabilizing deformations of $\GQR$, generated, \eg, by $\chi$, may
lead either towards the future (isomorphic to
$\mathfrak{so}(1,5)$) or
towards the elsewhere (isomorphic to $\mathfrak{so}(2,4)$). 
Both choices introduce non-commutativity of the $P$'s, 
differing in the sign of the associated curvature.%
}} 
\label{instPlot}
\end{figure}

A glance at Fig.~\ref{instPlot} helps visualize several
aspects of the stability analysis that were mentioned earlier
(see Sect.~\ref{OaH3}). For example, starting at some
(unstable)  point of, say, the future cone, it is clear that
infinitesimal deformations along the cocycle $\chi$ given above 
lead either to the future or to the elsewhere, both being
non-isomorphic to the original algebra but also between
themselves.  
Since $\llbracket \chi,\chi \rrbracket=0$, one may
also consider finite deformations, 
\ble{chifinite}
\mu_t =
\muPH(q,\alpha_1,\alpha_2,\sqrt{\alpha_1 \alpha_2})+t \chi
\, .
\ee
In this case,
it is clear that there exists a negative value of $t$
($t_0=-\alpha_1-\alpha_2$) such that the resulting algebra 
$\mu_{t_0}=\muPH(q,-\alpha_2,-\alpha_1,\sqrt{\alpha_1 \alpha_2})$ 
lies on the past light cone and is, therefore, non-isomorphic
to $\mu_t$, for generic $t$. Finally, had we chosen instead a
non-trivial cocycle orthogonal to the axis of the cone, rather
than parallel to it%
\footnote{%
This can achieved by adding an appropriate trivial cocycle to
$\chi$.%
}, 
there would exist a finite deformation
along it
isomorphic to the original algebra, given by the ``antipodal''
point on the future cone, 
$(\alpha_2,\alpha_1,-\sqrt{\alpha_1 \alpha_2})$.   
\subsection{Gauging away the \mvb $\alpha$'s \mvn} 
\label{Gaa3}
The above conclusion makes it evident that, for each of the
above classes, a representative exists with $\alpha_3=0$. The
deformation space then, from the algebraic point of view, 
is essentially the
$\alpha_1$-$\alpha_2$ plane. To find explicitly a linear
redefinition of the generators that moves an arbitrary point
in $\alpha$-space to the $\alpha_1$-$\alpha_2$ plane,
we notice that the transformation $(P,Z) \mapsto
(P',Z')$, given by 
\ble{PZMredef}  
P'_\mu = a P_\mu + b Z_\mu
\, ,
\qquad
\qquad
Z'_\mu= c P_\mu + d Z_\mu
\, ,
\qquad
\qquad ad-bc=1
\, ,
\ee
leaves the value of $q=1$ in the Heisenberg commutator
invariant, while, in $\alpha$-space, it induces the
transformation $\vec{\alpha} \mapsto \vec{\alpha}'=M
\vec{\alpha}$, where
\ble{Mdef}
M=\left(
\begin{array}{ccc}
a^2 & b^2 & 2ab
\\
c^2 & d^2 & 2cd
\\
ac & bd & ad+bc
\end{array}
\right)
\, .
\ee
It can be checked that, in the particular case where the 
$P$-$Z$ transformation is a
rotation by an angle $\theta$, 
$\left( 
\begin{array}{cc}
a & b\\
c & d
\end{array}
\right)
=
\left( 
\begin{array}{cc}
\cos \theta & \sin \theta\\
-\sin \theta  & \cos \theta
\end{array}
\right)
$,
the matrix $M$ that results from~(\ref{Mdef}) describes
a rotation in $\alpha$-space, around the axis of the cone, 
by an angle of
$2\theta$, counterclockwise as seen from the future. Such a
rotation clearly leaves all isomorphism classes invariant and
can be chosen so as to move any particular point to 
the $\alpha_1$-$\alpha_2$ plane. Similarly, for algebras in the
elsewhere, isomorphic algebras exist with either $\alpha_1$ or
$\alpha_2$ equal to zero --- this is not true for algebras in
the past or the future (the orbits of the latter under the
above rotation do not intersect the $\alpha_1$-$\alpha_3$ or
$\alpha_2$-$\alpha_3$ planes). Having said that, due
consideration should be given to the fact 
that physicists can be single-minded in what regards their preferred
set of  generators, \eg, working
with arbitrary linear combinations of momenta and positions
could be frowned on. 
If such preferences are given priority, one might
be forced to work with a version of the deformed algebra with
$\alpha_3 \neq 0$.
\subsection{Isomorphisms} 
\label{Iso}
This brings our stability analysis to a conclusion. The algebra
$\GPH(q,\vec{\alpha})$ we have arrived at
is given by Eqs.~(\ref{LLcr})--(\ref{LZcr}) and
(\ref{PZcomm})--(\ref{ZMcomm}),
with corresponding 2-cochain
\ble{muPHqaf}
\muPH(\vec{\alpha})
= 
\muCR + q \psiH 
+\alpha_1 \zeta_1
+\alpha_2 \zeta_2
+\alpha_3 \zeta_3
\, .
\ee
The novelties are that the $P$'s don't commute, the $Z$'s
don't commute, $M$ is no longer central, and an extra term
appears in the Heisenberg commutator. The nature of the three
deformation parameters $\alpha_i$ is discussed in Sect.~\ref{Tnotd},
after the
physical identification of the generators has been carried out. 

As it has been pointed out in~\cite{Khr.Lez:03,Vil:94}, the
above algebra, off the instability cone, is isomorphic to
some $\mathfrak{so}(m,6-m)$, where $m$ depends on the signs
of $\alpha_1$, $\alpha_2$ (taking $\alpha_3=0$). 
The isomorphism, given in~\cite{Vil:94}, 
is as follows: denote the generators of $\mathfrak{so}(m,6-m)$
by $\{ J_{\mu \nu},\, J_{\mu 4},J_{\mu 5}, J_{45} \}$ 
 --- their commutation relations are analogous to those of the
Lorentz group, Eq.~(\ref{LLcr}), with a metric
$\bar{g}$ that is taken diagonal, with entries $\pm 1$, and
coinciding with $g$ in the Lorentz sector. 
Assuming the identifications
\ble{PZMiden}
P_\mu= \sigma J_{\mu 4}
\, ,
\qquad
\qquad
Z_{\mu}=\tau J_{\mu 5}
\, ,
\qquad
\qquad
M=\rho J_{45}
\, ,
\ee
one finds the commutators
\ble{PPZZMcomm}
[P_\mu,Z_\nu]=-i \, \frac{\sigma \tau}{\rho} \bar{g}_{\mu \nu}
M
\, ,
\qquad
[P_\mu,M]=i \, \frac{\rho \sigma}{\tau} \bar{g}_{44} Z_\mu
\, ,
\qquad
[Z_\mu,M]=-i \, \frac{\rho \tau}{\sigma} \bar{g}_{55} P_\mu
\, .
\ee
Comparing the first of these with the Heisenberg commutator
gives $\rho=-q\sigma \tau$. Substituting this in the other two, 
and comparing
with~(\ref{PMcomm}), (\ref{ZMcomm}), respectively, shows that 
\ble{bgdef}
\bar{g}=\text{diag}(1,-1,-1,-1,\epsilon_4,\epsilon_5)
\, ,
\qquad
\epsilon_4 \equiv
-\text{sgn}(q\alpha_1)
\, ,
\qquad
\epsilon_5 \equiv
-\text{sgn}(q\alpha_2)
\, ,
\ee
\ie, assuming $q>0$,
\ble{Giso}
\GPH(q,\alpha_1,\alpha_2, \alpha_3=0)
\cong
\begin{cases}
\mathfrak{so}(1,5)&
\text{if $\alpha_1 >0, \, \alpha_2 >0$}
\\
\mathfrak{so}(2,4)&
\text{if $\alpha_1 \alpha_2 <0$}
\\
\mathfrak{so}(3,3)&
\text{if $\alpha_1 <0, \, \alpha_2 <0$}
\end{cases}
\, .
\ee
(see Fig.~\ref{instPlot}). On the light-cone, the above
semisimple (and, hence, stable) algebras go over to the
corresponding semidirect product~\cite{Khr.Lez:03}.
\subsection{Relation with other algebras}
\label{Rwoa}
As mentioned in the Introduction, several non-commutative
spacetime Lie algebras have been proposed over the years, but, as
a rule, they fail to provide a complete set of generators. We
provide below an account of these earlier attempts,
emphasizing from the outset that our list of references does
not pretend to be complete. 

The first, to our knowledge, publication regarding a non-commutative,
Lorentz covariant spacetime is due to
Snyder, dating from 1947. Apparently, as J.{}
Wess has documented and publicized, the idea can be traced
back to Heisenberg, who, in a letter to Peierls, suggested
that the ultraviolet infinities of quantum field theory could
be tamed by assuming noncommuting spacetime coordinates.
Peierls soon found an altogether different application in the 
calculation of the
lowest Landau level of a system of electrons in a magnetic
field with impurities, using noncommuting coordinates in the
potential-like function describing them. He also passed on the 
idea to Pauli,
who described it to Oppenheimer, who shared it with Snyder,
who published Ref.~\cite{Sny:47} (our source
is~\cite{Jac:01}). Snyder's position operators
fail to commute among themselves, exactly as in~(\ref{ZZcomm}). 
His momenta, however, commute, and the
position-momenta relations contain non-linear terms. An early
attempt at formulating electrodynamics in this non-commutative 
spacetime
followed shortly after~\cite{Sny:47a}. Later
in that year, Yang~\cite{Yan:47}, pointed out
that by introducing what we have called $M$, one can render
the algebra linear. Additionally, he proposed non-commuting
momenta, exactly as in~(\ref{PPcomm}) {\em and}
the accompanying $P$-$M$ relations, Eq.~(\ref{PMcomm}) (with
$\alpha_3=0$).
The $\mathfrak{so}$ isomorphism
was also given, although with a particular choice
for the signs $\epsilon_4$, $\epsilon_5$ (both equal to $-1$). 
In fact, getting to
some known, preferably semisimple, algebra (like
$\mathfrak{so}$), seems to have been his guiding
principle in completing the set of commutators. 

Several years
later, Khruschev and Leznov~\cite{Khr.Lez:03} provided a
further deformation, essentially the one given by our $\zeta_3$,
\ie, by the terms proportional to $\alpha_3$ in
(\ref{PZcomm})--(\ref{ZMcomm}). Although their article
cited above appeared in 2002, it quotes this result (or, at
least, significant parts of it) from an earlier work
of theirs, dating from 1973, which we have not had access to.
Their approach is via a straightforward solution of the Jacobi
identities, and does not include the stability point of
view. 
Apart from the deformation itself, they also provide
information on the Casimirs of the deformed algebras, and
mention the $\alpha_1 \alpha_2 -\alpha_3^2 \neq 0$ relation, with
$\alpha_1 \alpha_2 \neq 0$, as a
semisimplicity criterion for the deformed algebra. 
The $\mathfrak{so}$ identifications are given (they actually
use $\mathfrak{o}(n,6-n)$) but the 
possibility of gauging away $\alpha_3$ by a redefinition
of the generators is not pointed out. Some steps towards
constructing field theories on these quantized spaces were taken
in~\cite{Lez:04}.

The work of Vilela
Mendes~\cite{Vil:94}, which motivated ours, appeared in 1994.
There, for the first time, the stability criterion for the
non-commutative spacetime algebra is invoked
and its relevance, more generally, for the algebraic structures 
employed in physical theories is convincingly advocated. The
approach taken in determining the stable form of the algebra
is a minimalistic one: the $\mathfrak{so}$ algebras
are proposed {\em ab initio}, being obviously
deformations, and their semisimplicity is
invoked to guarantee their stability. 
Economical as it may be
this approach, it leaves neverheless pending the question of 
uniqueness, prompting us to undertake the present systematic
search. The instability double cone is not mentioned in the
above work. Also,
although Snyder's work was known to the author, it seems he
did not come across Yang's contribution. 
Various applications 
have been considered by Vilela Mendes and co-workers 
in~\cite{Car.Vil:01,Vil:04,Vil:00}.

In recent years, as mentioned in the Introduction, several 
``Doubly Special Relativity'' algebras
have been proposed. They all ignore the (initially) central
generator $M$. In~\cite{Kow.Now:02} it was shown that all
commutation relations of the deformed algebras, except the 
$P$-$Z$ ones, can be brought
into a Lie form by appropriate non-linear redefinitions of the
generators. The Lie form found coincides with that provided by
the $\alpha_2$ deformation above. Furthermore, it was pointed out
in~\cite{Chr.Oko:04a}, that by taking $M$ into account, the
``Triply Special Relativity'' of~\cite{Kow.Smo:04} is
linearized, and the resulting, Lie-type, deformation is the one provided
by $\alpha_1$ above (this was essentially a repeat of Yang's 
observation on Snyder's proposal, applied to the momentum sector). 
Thus, it seems that when $M$ is taken into account,
non-linear redefinitions bring the ``Multi-Special Relativity''
algebras into one of the forms found above%
\footnote{%
This statement is meant as an observation of a pattern, not as
a theorem --- we have certainly not checked each and every
non-linear algebra proposed.%
}. These observations suggest that, before leaving the tried
and tested Lie algebra framework, Lie deformations introducing
new invariant scales, like the ones proposed earlier, 
should be studied carefully, and the need
for non-linearity should be critically examined. 

Finally, the following obvious fact should be emphasized: 
isomorphic, or even identical,
algebras may correspond to radically different physics if the
generators that enter in them are interpreted in different
ways. In this respect, all of the above mentioned works
coincide in the physical identification of the generators, in
particular, in the fact that the $Z_\mu$'s should be
interpreted as position
operators. As we explain in the section that follows, our view differs.
\section{Some Physical Considerations}
\label{SPC}
We deal, finally, with a number of interpretational issues. 
We would like to warn the reader that the material in this
section is still in its formative stage, and several aspects
of what follows are still under investigation.
Nevertheless, we feel it is worthwhile pointing out
alternative possibilities in the physical identification of
the generators we have been studying. 
The content here is mostly 
qualitative and the tone, accordingly, informal. 
We keep
complexity to a minimum by treating the case of a massive,
spinless particle only ---  
a more complete analysis will have to wait,
like so many other things, a future work. 
\subsection{The coproduct of Lie algebra generators}
\label{TcoLag}
We wish to discuss the physical meaning of the coproduct 
of Lie algebra generators. We will use the Poincar\'e algebra
as an example, but the discussion applies to
general Lie algebras. 

Consider applying a translation $T_{\vec{a}}$ to a
particle, located at $\vec{x}$. As a result, the particle
shifts to $\vec{x}+\vec{a}$. 
Imagine now that, under closer
inspection, the particle is seen to be a bound state of two
other particles. To translate by $\vec{a}$ what is now known
to be a two-particle system, one applies the translation
$T_{\vec{a}}$ to each of the constituent particles in the
system. The $n$-particle case, $n>2$, is handled by further
subdivision of either of the two particles above. Similar
considerations hold for rotations or boosts. This observation
is formalized in the following manner. The state of the system
under study is represented by a state vector $\ket{\psi}$ in
some Hilbert space $\calH$. To a possible transformation 
of the system,
\eg, a rotation $R_{\alpha\beta\gamma}$ parametrized by
Euler's angles, one associates an operator
$\calD(R_{\alpha\beta\gamma})$, acting on $\calH$. When the
system is revealed to consist of, say, particles 1 and 2, the state
space becomes $\calH_1 \otimes \calH_2$, where $\calH_i$ is
the state space of particle $i$. The observation made above
then implies that the operator representing
$R_{\alpha\beta\gamma}$ in $\calH_1 \otimes \calH_2$ is
simply $\calD_1(R_{\alpha\beta\gamma}) \otimes
\calD_2(R_{\alpha\beta\gamma})$, where $\calD_i$ is the
representation of rotations in $\calH_i$. This is true for all
representations $\calD_i$ --- we may accordingly
conclude that the abstract rotation operator
$R_{\alpha\beta\gamma}$ acts on tensor products as
$R_{\alpha\beta\gamma} \otimes R_{\alpha\beta\gamma}$ and
call this latter operator the {\em coproduct} $\Delta(R_{\alpha
\beta\gamma})$ of $R_{\alpha\beta\gamma}$. Particular cases
then are handled by taking the appropriate representation of
this universal formula, \eg, $\calD_1 \otimes \calD_2$ above.
The fact that rotations should compose in the same way,
whether applied to a simple or to a composite system, is
expressed algebraically by the requirement that%
\footnote{%
\label{fn:Dhomo}
Another way of writing this is
$[\Delta(T_A),\Delta(T_B)]=f_{AB}^{\phantom{AB}C}
\Delta(T_C)$, \ie, the generators $\Delta(T_A)$ in $\calG
\otimes \calG$ satisfy the same commutation relations as the
$T_A$'s.%
} 
$\Delta(R_1
R_2) = \Delta(R_1) \Delta(R_2)$. Our formalism respects this
for, if $R_1 R_2=R_3$, then
\bae
\label{Dhomo}
\Delta(R_1 R_2)
\fe
\Delta(R_3)
\ff
 \fe
R_3 \otimes R_3
\ff
 \fe
R_1 R_2 \otimes R_1 R_2
\ff 
\fe
(R_1 \otimes R_1)(R_2 \otimes R_2)
\ff
 \fe
\Delta(R_1) \Delta(R_2)
\, ,
\eae
the product law in the tensor product being $(A \otimes B)(C
\otimes D) = AC \otimes BD$. 
We summarize: for all transformations $T$ in the Poincar\'e
group
\begin{itemize}
\item
the coproduct $\Delta(T)$ is {\em grouplike},
\ble{copglike}
\Delta(T) = T \otimes T
\ee
\item
$\Delta$ is an algebra homomorphism,
\ble{Dhomo2}
\Delta(T_1 T_2) = \Delta(T_1) \Delta(T_2)
\, .
\ee
\end{itemize}
Now write $T=e^A$, with $A$ in the Poincar\'e algebra
$\GP$ and
{\em define} $\Delta$ to be linear in the entire $U(\GP)$, 
the universal enveloping algebra%
\footnote{%
More precisely, a certain topological completion of
$U(\GP)$.%
} of $\GP$ ---
a simple calculation then shows that $\Delta(A)= A \otimes 1 +
1 \otimes A$ (this is a logarithm turning a product into a sum,
as usual). We conclude that
\begin{itemize}
\item
The generators of grouplike transformations are
{\em primitive},
\ble{primidef}
\Delta(A)= A \otimes 1 + 1 \otimes A
\, ,
\ee
\end{itemize}
with $J_{\text{tot}}=J_1 + J_2$ as the archetypical example
from quantum mechanics. In other words, {\em the physical quantities
corresponding to generators of grouplike transformations are
additive under system composition} (or {\em extensive}, in 
thermodynamics parlance). All Lie algebra generators 
are of this nature. From a geometric point of view, the
definition of pointwise multiplication of functions on the
group manifold, $(fh)(g)=f(g)h(g)$, is what fixes the coproduct of the
point (transformation) $g$ to be grouplike. 
At the infinitesimal level, this
becomes the Leibniz rule satisfied by the generators (this is
another way of interpreting~(\ref{primidef})), consistent with 
their representation as first-order differential
operators on the group manifold. The above considerations
prompt us to only allow primitive operators as Lie algebra
generators. 

In~\cite{Vil:94}, it is argued that the 1-dimensional
Heisenberg commutator, $[p,x]=-i$, can be interpreted as
defining a stable Lie algebra. The justification
for this claim is made through the observation that one could
equally well choose a function of $x$ as a coordinate, in
particular, $y=e^{ix}$. Then, the Heisenberg relation takes
the form $[p,y]=y$, which indeed defines a stable
2-dimensional algebra. Apart from the unsuitability of
$e^{ix}$ as coordinate over the entire $x$-axis, it should be 
clear from our earlier 
discussion that we cannot agree with this argument, since $y$
is no more primitive than $x$.
\subsection{The Lie form of the Heisenberg algebra}
\label{PwtHa}
As mentioned after our first reference to the Heisenberg
commutator, Eq.(\ref{HeisenLie}), there are a number of
remarks that we would like to make regarding its proposed form.
One usually first encounters the Heisenberg commutator in the
form
\ble{Heisen1}
[P_i,X_j]=-i \, q \delta_{ij}
\, ,
\ee
which is unsatisfactory for (at least) two reasons.
The first has to do with Lorentz covariance --- the obvious
remedy is to consider instead the form 
\ble{Heisen2}
[P_{\mu},X_{\nu}]=i \, q  g_{\mu \nu}
\, ,
\ee
leaving for a future brainstorm the elucidation of its
physical implications (notice that time is promoted to an
operator). 
The second reason is of a technical nature: 
dealing with a Lie algebra,
the \rhs{} of~(\ref{Heisen2}) ought to be linear in the
generators. The usual solution followed in the literature is to
introduce a new, central generator $M$, with 
\ble{Heisen3}
[P_{\mu}, X_{\nu}]= i \, q g_{\mu \nu} M
\, .
\ee 
The resulting three-generator Lie algebra is referred to as the 
Heisenberg algebra --- the physical interpretation of 
$M$ is generally left obscure. 
It might at
first seem that there is little to be gained from writing out
$M$ explicitly, since it commutes with everything, but when 
deformations of the algebra are
considered, it will be essential to do so since, as a result
of the deformation, $M$ might cease to be central
(for an example of the type of problems that may arise by
suppressing $M$, see~\cite{Chr.Oko:04a}). 

Is~(\ref{Heisen3}), at last, an acceptable form of the
Heisenberg algebra? That the answer should still be negative follows 
easily from our remarks about the primitiveness of Lie algebra
generators. First, if $M$ in the \rhs{} of~(\ref{Heisen3})
were primitive (and hence extensive), the effective Planck's
constant for a composite system would be the sum of those for
its constituent parts, providing for several concrete examples
of fuzzy spheres (\eg, the earth, with 
$q_{\text{Earth}} \cong 10^{14}$ Kgr m$^2$/sec). Second,
$X_{\mu}$ is not primitive. There are various
ways to see this. To begin with, it is rather obvious that position is
not an extensive quantity: if two particles are glued
together at $x_{\mu}$, their composite system is also located
at $x_{\mu}$, not at $2x_{\mu}$. Another way is to look at the
corresponding finite transformation. $X_{\mu}$ may be
considered, up to a sign, the generator of translations in 
momentum space. But the apparent symmetry (via duality) between 
momenta and positions should be treated with care. 
In particular, although translations in spacetime are grouplike, 
those in momentum space are not. If
a particle of 4-momentum $p$ is translated, in momentum space,
by $k$, it ends up with 4-momentum $p+k$. If now it is
discovered that it is actually made up of two other particles
and each of them is translated by $k$, then the composite
particle would be translated by $2k$. This latter example
reveals something about the nature of the grouplike operator 
that should replace $e^X$, the logarithm of which would be
acceptable as a Lie algebra generator. 
Roughly speaking, it should somehow detect 
the mass of the
particle and translate in momentum space by a quantity
proportional to it. Notice that, despite the elementary nature of the
considerations in this section, there seems to exist a
consensus in the literature that $X_{\mu}$ {\em is} primitive%
\footnote{%
The references assuming so are too many to list
here explicitly ---~\cite{Kow.Now:02} may nevertheless be
singled out for actually {\em deriving} this result
(see their Eq.~(26) and the erroneous argument 
preceding it).%
}.
\subsection{The coproduct of the position operator}
\label{Tcotpo}
So, if $X_{\mu}$ is not primitive, what is its coproduct
$\Delta(X_{\mu})$? The
answer is that, in general, $\Delta(X_{\mu})$ does not exist. 
To see why this is so, let us first specify what exactly is it that
we want the position operator to do for us. For a single localized 
particle
it is clear that $X_{\mu}$ should return its position, but
what should $X_{\mu}$ (via its coproduct $\Delta(X_{\mu})$) 
do on a two-particle system? 
Clearly, if the two particles are glued together and the
composite system is localized, we should get the same answer 
whether we operate with $X_{\mu}$ on the composite particle or 
with $\Delta(X_{\mu})$ on the two-particle system. 
When the two particles are far
apart and/or have different velocities, the natural
requirement would be that $\Delta(X_i)$ (\ie, the spatial part of
$\Delta(X_{\mu})$) should return the position
of their center-of-momentum (or center-of-inertia), 
\ie, the relativistic refinement
of the newtonian center-of-mass concept, which is the natural
``effective position'' of a relativistic composite system%
\footnote{%
See, for example, the discussion in~\cite{Rin:79}, p.~84
and~\cite{Lan.Lif:83}, p.~42.%
}. The problem is that
the center-of-momentum 3-vector is not the spatial part of any
4-vector, in other words, the ``effective position'' of a
composite relativistic system does not behave as a 4-vector.
As a result, different observers locate the center-of-momentum
of a system at different points. 
This, in turn, implies that $\Delta(X_{\mu})$
does not satisfy the same commutation relations
with $\Delta(L_{\rho \sigma})$ as $X_{\mu}$
does with $L_{\rho \sigma}$ (see footnote~\ref{fn:Dhomo}), in other 
words, $\Delta$ fails to
be a homomorphism of the algebra, which proves our assertion. 

The above conclusion might well be correct, but, certainly,
there are composite systems the ``effective position'' of
which, for all practical purposes,
behaves like a 4-vector (\eg, an $\alpha$-particle). This implies 
that although, strictly
speaking, $\Delta(X_{\mu})$ does not exist, one might
nevertheless define an approximate coproduct that works
provided it is applied on a restricted class of systems --- 
intuitively, systems that can fool the observer into thinking
they are a single, localized particle. To make
this statement precise, we note that the center-of-momentum
spatial coordinates of a (non-interacting) two-particle system 
are given by
\ble{com2ps}
\vec{R}=\frac{E_1 \vec{r}_1 + E_2 \vec{r}_2}{E}
\, ,
\ee
where $E \equiv E_1+E_2$ is the total energy of the system (this
formula makes it clear that $\vec{R}$ is not the spatial part of
any 4-vector). Assume now%
\footnote{%
Our simplifying assumptions of non-zero mass and zero spin
start taking effect from this point on.%
} that the system under study is such
that in its center-of-momentum frame all
energies $E_i$ are nearly equal to the corresponding rest masses, 
$E_i \cong m_i$ --- we will call such a system {\em psychron},
from the greek \selg yuqr\'on \sele for ``cold''. 
Then, in the above frame, (\ref{com2ps}) reduces to the Newtonian
formula for the center-of-mass. Moreover, when boosting to an 
arbitrary frame,
all energies in the \rhs{} of~(\ref{com2ps}) transform by the
same $\gamma$-factor, which cancels, so that the \lhs{}
transforms as a spatial vector.  We conclude that,
for psychron 2-particle systems, the relation
\ble{Xmudef}
m_{12} x^{\mu}_{12}= m_1 x^{\mu}_1 + m_2 x^{\mu}_2
\, ,
\ee
where $m_{12} \equiv m_1 + m_2$, defines the effective position
$x_{12}$ of the system as a 4-vector.
Denoting by $M$ the mass operator, $M^2=P^{\mu}P_{\mu}$, and
brushing aside ordering ambiguities, we
conclude from~(\ref{Xmudef}) that the {\em moment operator}
$Z_{\mu} \equiv X_{\mu} M$ {\em is} primitive, 
when applied to psychron systems ($M$ is also primitive on such
systems). We note
furthermore that $Z_{\mu}$ is of exactly the form anticipated
by the argument at the end of Sec.~\ref{PwtHa}.
In terms of
$Z$, the covariant version of the Heisenberg relation,
Eq.~(\ref{Heisen2}), takes the familiar form used earlier,
\be
\tag{\ref{HeisenLie}$'$} 
[P_{\mu},Z_{\nu}]=i \, q g_{\mu\nu} M
\, ,
\ee
albeit with a new interpretation.
\subsection{The algebra of standard quantum relativistic
kinematics}
\label{Taosqrk}
We investigate the repercussions of the above interpretation
of $Z_\mu$, $M$, in identifying the algebra of standard quantum
relativistic kinematics. In the latter, the momenta commute
and so do the positions, while their cross-relations are given by the
Heisenberg commutator, Eq.~(\ref{Heisen2}).   
But then the $Z$'s, in terms of which the algebra should be
expressed, do not commute,
\ble{ZZnoncomm}
[Z_\mu,Z_\nu]= i \, q (X_\mu P_\nu -X_\nu P_\mu)
\, ,
\ee
and neither do the $Z$'s with $M$,
\ble{ZMnoncomm}
[Z_\mu, M]=-i \, q P_\mu
\, ,
\ee
where $[X_\mu,f(P)] = -i \, q \partial f(P)/\partial P^\mu$
was used%
\footnote{%
Strictly speaking, this relation holds for functions $f(P)$
that can be expanded in power series in $P$ ---
nevertheless, the commutation 
relations~(\ref{ZZnoncomm}), (\ref{ZMnoncomm}), are consistent 
with $M^2=P^\mu P_\mu$ and we do not require anything more.%
}. 
Notice that the $Z$-$Z$ non-commutativity  is a purely quantum 
($q \neq 0$) phenomenon and has
no connection to spacetime non-commutativity.
We recognize the \rhs{} of~(\ref{ZZnoncomm}) as (a multiple of) 
the covariant form of the orbital 
angular momentum generator, $L_{\mu \nu}=q^{-1}(X_\mu P_\nu -X_\nu
P_\mu)$. For a massive, spinless particle then, we have
\ble{ZZcomm2}
[Z_\mu,Z_\nu]=i \, q^2 J_{\mu \nu}
\, .
\ee
A look at~(\ref{ZZcomm}), (\ref{ZMcomm}), shows that the above 
relations, 
Eqs.~(\ref{ZMnoncomm}) and (\ref{ZZcomm2}), are of
exactly the form furnished by the $\alpha_2$ deformation, with
$\alpha_2=q$. We conclude that, for a massive, spinless
particle, the algebra $\GQR$ of standard quantum relativistic
kinematics is given by 
\ble{GQR}
\GQR=\GPH(q,0,q,0)
\, ,
\ee
\ie, in $\alpha$-space, it lies on
the surface of the future cone, along the $\alpha_2$-axis, at
$\alpha_2=q$ (see Fig.~\ref{instPlot}). 
It is worth noting that, with the above interpretation of the
$Z$'s, the Heisenberg algebra by itself does not close, the
$Z$-$Z$ commutator generating the Lorentz group. 
\subsection{The nature of the deformations}
\label{Tnotd}
The deformation corresponding to $\zeta_1$ introduces 
non-commutativity among
the momenta and renders $M$ non-central. Its origins lie in
the instability of the Poincar\'e algebra,
which stabilizes to the simple De Sitter algebras
$\mathfrak{so}(1,4)$ or $\mathfrak{so}(2,3)$. The corresponding
parameter, $\alpha_1$, has dimensions $[L]^{-1}[M]$, so that
$R \equiv \sqrt{\hbar/\alpha_1}$ is a length, the radius of
curvature of the manifold on which the various $\mathfrak{so}$
algebras of Sect.~\ref{Iso} act. It has been suggested
in~\cite{Vil:94} that, as long as one is interested in the
kinematics in the tangent space to the manifold, rather than
the group of motions of the manifold itself, one may take
the $R\rightarrow \infty$ limit, \ie, one may essentially 
disregard the
above deformation. On the other hand, in~\cite{Kow.Smo:04},
the suggestion has been made that $R^2$ may set the scale for
the cosmological constant $\Lambda$. In any case, this
deformation is a familiar and thoroughly studied one.

When the $Z_\mu$ are identified with the position operators, 
the deformation generated by $\zeta_2$ turns on spacetime
non-commutativity. $\alpha_2$ in that case has dimensions
$[L][M]^{-1}$ , so that $\ell \equiv
\sqrt{\hbar \alpha_2}$ is a length, the inverse of which 
has, in the past, been conjectured to set the scale for
the masses of the elementary particles. However, if that were
the case, the effects of the deformed commutators would by now
have been measured, so this proposal had to be abandoned. A more
recent tendency is to regard $\ell$ as the Planck length, and
attribute the non-commutativity to quantum gravity effects
(see, \eg,~\cite{Kow.Smo:04} and references therein). 
Whatever the interpretation of
the new length scale may be, the above identification of the
$Z$'s seems to us to suffer from a somewhat incredulous
prediction: the extent to which the coordinates of a particle
do not commute, \ie, the local ``fuzzines'' in spacetime due to,
\eg, quantum gravity effects, depends, in general, on the position 
of the origin
(since $J_{\mu \nu}$, the particle's angular momentum, does). 
In particular, the coordinates of a
particle at the origin commute. We think it improbable
that such a state of affairs can be succesfully incorporated in
a consistent physical scheme, and invite workers pursuing this
direction to address what, to us, seems like a neglected pathology. 
In conclusion, then, we think it fair to say that interpreting
the $Z$'s as spacetime coordinate operators of a particle 
makes it improbable for the
$\alpha_2$ deformation to have the physical applications proposed 
in the literature. On the
other hand, our identification of the $Z$'s with the moment
operators leads to the conclusion that $\GQR$, the
standard, experimentally tested, quantum relativistic algebra
in which, in particular, 
the spacetime coordinates commute, 
is given, in the case of a massive spinless particle,  
by $\alpha_2=q$, with  the experiment fixing the value
$q=\hbar$. If the interpretation advocated above is
correct, then, a look at Fig.~\ref{instPlot} shows that the
only deformations left to explore are those generated by
$\pm \chi$, leading to the future or the elsewhere,
respectively, both introducing non-commutativity of the
momenta. 

The $\alpha_3$ deformation signals a more radical departure
from $\GQR$, so much so that, in Ref.~\cite{Vil:94}, it is 
practically discarded as
unphysical. Ref.~\cite{Khr.Lez:03}, on the other hand, treats
it on an equal footing and observes that, with the $Z$'s as
positions, $\alpha_3$ is dimensionless, so that $\hbar
\alpha_3$ is a new fundamental constant with dimensions of
action. When the $Z$'s are taken as moments, $\alpha_3$
acquires dimensions of mass. In either case, the physical
implications of the deformation are somewhat obscure and
deserve further study. 
\section{Concluding Remarks}
\label{CR}
We have pursued in this paper the stability point of view to
its ultimate consequences. Our systematic algebraic analysis has
recovered previous results, establishing their
uniqueness, and shedding light along
the way on various technical issues, in particular, the
interrelations among the deformations found. A
fundamental departure from the established lore has been our
identification of the $Z_\mu$ generators with the moment operators
of a (massive, spinless) particle, having concluded that the
position operators lack the essential property of
primitiveness, necessary for all Lie algebra generators. 

We think  that a number of questions raised here deserve further
study. First, we would like to generalize the concept of the moment
operators to the case of particles with spin, and/or zero
mass. Second, representation theoretical aspects of the
problem should be examined, in particular, a Wigner-type 
classification should be carried through. 
It would also be of interest to develop some
degree of intuition regarding the deformed kinematics, \eg, by
clarifying the coexistence of the Lorentz contraction with 
an invariant length scale.
\section*{Acknowledgements}
\label{Ack}
One of the authors (C.C.) acknowledges partial financial support 
from DGAPA-PAPIIT grants IN
114302, IN 108103-3 and CONACyT grant 41208-F.
\appendix
\section{Computing \mvb $H^2(\GCR)$ \mvn}
\label{H2GCR}
The complexity of the
calculation of
the second cohomology group of an algebra 
grows rapidly with its dimension. When dealing 
with a fifteen-dimensional algebra, like $\GCR$ in the case at hand, 
the prospect of carrying out the analysis manually becomes 
somewhat unattractive. Luckily, some \textit{MATHEMATICA} code 
we wrote deals with the problem within minutes ---
we give here some details of the calculations. The algorithm
we used was the following:
\begin{enumerate}
\item
Consider the most 
general 1-cochain $\phi$,
\be
\phi= \phi_{A}^{\phantom{A}B} \Pi^A \otimes T_B
\, ,
\ee
with $\phi_{A}^{\phantom{A}B}$ arbitrary real constants 
(a sum of $15^2=225$ terms). Obtain
the most general 2-coboundary $\psi$ by setting $\psi= \nabla \phi$. 
This produces a sum of 
1008 terms, each corresponding to a non-zero 
component of $\psi$.
\item
Consider the most general 2-cochain $\chi$,
\be
\chi= \chi_{AB}^{\phantom{AB}C} \Pi^A \Pi^B \otimes T_C
\, ,
\ee
with $\chi_{AB}^{\phantom{AB}C}$ arbitrary real constants 
(a sum of $15 \binom{15}{2}=1575$ terms). Require that it 
be a 2-cocycle by setting $\nabla \chi=0$. This results 
in a system of $5672$ linear homogeneous equations in the 
above 1575 $\chi_{AB}^{\phantom{AB}C}$'s, which is solved
for some of them in terms of the 
rest --- call the latter $c_i$. Effecting these substitutions
in $\chi$, one obtains the most general 2-cocycle  
$\tilde{\chi} \equiv \sum_i c_i \chi_i$ with {\em arbitrary} $c_i$. 
As a result, 
each of the $\chi_i$ in the sum is by itself a 
2-cocycle --- there are 221 of them in our case.
\item
Examine which of the $\chi_i$'s  
are non-trivial, \ie, check if the equations 
$\chi_i = \psi$ have a solution for the
$\phi_A^{\phantom{A}B}$ that appear in $\psi$. 
For each $\chi_i$, this produces a system of 1575 equations. 
If a solution exists, the 2-cocycle in question is trivial,
\ie, a 2-coboundary. For the problem at hand, 5 out of the 
211 $\chi_i$ turn out to be non-trivial. 
\item
Check whether the non-trivial 
cocycles obtained correspond to {\em independent} generators of 
$H^2(\GCR)$. Do this by setting an arbitrary linear
combination of the 2-cocycles equal to the general 2-coboundary. 
If a solution for 
the $\phi_{A}^{\phantom{A}B}$ exists, discard one of the 
cocycles that enter in the linear combination and repeat the test 
for the  remaining ones, until no solution exists. 
For the case at hand, no linear dependence was found, 
arriving thus at the final result, Eq.~(\ref{H2GCRdef}).
\end{enumerate}

\begin{thebibliography}{10}

\bibitem{Ahl:03}
D.{}~V.{} Ahluwalia-Khalilova.
\newblock Operational {I}ndistinguishability of {D}oubly {S}pecial
  {R}elativities from {S}pecial {R}elativity.
\newblock \texttt{gr-qc/0212128} (version 2, June 2003).

\bibitem{Ame:02a}
G.~Amelino-Camelia.
\newblock Relativity in {S}pacetimes with {S}hort {D}istance 
{S}tructure
  {G}overned by an {O}bserver {I}ndependent ({P}lanckian) 
{L}ength {S}cale.
\newblock {\em Int.{} J.{} Mod.{} Phys.{} D}, 11:35--60, 2002.

\bibitem{Anc:88}
J.{} M.{}~Ancochea Bermudez.
\newblock On the {R}igidity of {S}olvable {L}ie {A}lgebras.
\newblock In M.{} Hazewinkel and M.{} Gerstenhaber, 
editors, {\em Deformation
  {T}heory of {A}lgebras and {S}tructures and 
{A}pplications}, pages 403--445.
  Kluwer Academic Publishers, 1988.

\bibitem{Car.Vil:01}
E.{} Carlen and R.{}~Vilela Mendes.
\newblock Non-commutative {S}pace-time and the 
{U}ncertainty {P}rinciple.
\newblock {\em Phys.{} Lett.{} A}, 290:109--114, 2001.

\bibitem{Che.Eil:48}
C.~Chevalley and S.~Eilenberg.
\newblock Cohomology {T}heory of {L}ie {G}roups and {L}ie {A}lgebras.
\newblock {\em Trans. Am. Math. Soc.}, 63:85--124, 1948.

\bibitem{Chr:01a}
C.{} Chryssomalakos.
\newblock Lie {S}uperalgebra {S}tability and {B}ranes.
\newblock {\em Mod.{} Phys.{} Lett.{} A}, 16:197--210, 2001.
\newblock \texttt{hep-th/0102134}.

\bibitem{Chr.Azc.Mac.Bue:99}
C.~Chryssomalakos, J.{}~A.{} de~Azc\'arraga, 
A.{}~J.{} Macfarlane, and J.{}
  C.{}~Perez Bueno.
\newblock Higher {O}rder {BRST} and {A}nti-{BRST} {O}perators 
and {C}ohomology for {C}ompact {L}ie {A}lgebras.
\newblock {\em J.{} Math.{} Phys.}, 40/10:6009--6032, 1999.

\bibitem{Chr.Oko:04a}
C.{} Chryssomalakos and E.{} Okon.
\newblock Linear {F}orm of 3-scale {R}elativity {A}lgebra 
and the {R}elevance of {S}tability.
\newblock {\em Int.{} J.{} Mod.{} Phys.{} D}, 2004.
\newblock \texttt{hep-th/0407080}.

\bibitem{Azc.Izq:95}
J.{}~A.{} de~Azc\'arraga and J.{}~M.{} Izquierdo.
\newblock {\em Lie {G}roups, {L}ie {A}lgebras, {C}ohomology and {S}ome
  {A}pplications in {P}hysics}.
\newblock Cambridge University Press, 1995.

\bibitem{Ger:64}
M.~Gerstenhaber.
\newblock On the {D}eformation of {R}ings and {A}lgebras.
\newblock {\em Ann. Math.}, 79:59--103, 1964.

\bibitem{Ger.Sch:88}
M.{} Gerstenhaber and S.{}~D.{} Schack.
\newblock Algebraic {C}ohomology and {D}eformation {T}heory.
\newblock In M.{} Hazewinkel and M.{} Gerstenhaber, editors, 
{\em Deformation
  {T}heory of {A}lgebras and {S}tructures and {A}pplications}, 
pages 1--264.
  Kluwer Academic Publishers, 1988.

\bibitem{Ger.Sch:90}
M.~Gerstenhaber and S.{}~D.{} Schack.
\newblock Bialgebra {C}ohomology, {D}eformations, and 
{Q}uantum {G}roups.
\newblock {\em Proc.{} Nat.{} Acad.{} Sci.{} U.S.A.}, 87:478--481, 1990.

\bibitem{Goz:88}
M.{} Goze.
\newblock Perturbations of {L}ie {A}lgebra {S}tructures.
\newblock In M.{} Hazewinkel and M.{} Gerstenhaber, 
editors, {\em Deformation
  {T}heory of {A}lgebras and {S}tructures and {A}pplications}, 
pages 265--355.
  Kluwer Academic Publishers, 1988.

\bibitem{Hoc.Ser:53}
G.{} Hochschild and J-P.{} Serre.
\newblock Cohomology of {L}ie {A}lgebras.
\newblock {\em Ann.{} Math.}, 57/3:591--603, 1953.

\bibitem{Jac:01}
R.{} Jackiw.
\newblock Physical {I}nstances of {N}oncommuting {C}oordinates.
\newblock 2001.
\newblock \texttt{hep-th/0110057}.

\bibitem{Jac:79}
N.~Jacobson.
\newblock {\em Lie Algebras}.
\newblock Dover Publications, New York, 1979.

\bibitem{Khr.Lez:03}
V.{}~V.{} Khruschev and A.{}~N.{} Leznov.
\newblock Relativistic {I}nvariant {L}ie {A}lgebras for {K}inematical
  {O}bservables in {Q}uantum {S}pacetime.
\newblock {\em Grav.{} Cosmol.}, 9:159, 2003.
\newblock \texttt{hep-th/0207082}, v.{} 5 (March 2003).

\bibitem{Kow.Now:02}
J.{} Kowalski-Glikman and S.{} Nowak.
\newblock Non-commutative {S}pace-{T}ime of {D}oubly {S}pecial 
{R}elativity
  {T}heories.
\newblock 2002.
\newblock \texttt{hep-th/0204245}.

\bibitem{Kow.Smo:04}
J.{} Kowalski-Glikman and L.{} Smolin.
\newblock Triply {S}pecial {R}elativity.
\newblock 2004.
\newblock \texttt{hep-th/0406276}.

\bibitem{Lan.Lif:83}
L.{}~D.{} Landau and E.{}~M.{} Lifshitz.
\newblock {\em The {C}lassical {T}heory of {F}ields}.
\newblock Pergamon Press, 1983.
\newblock (Fourth edition, reprinted in 1983).

\bibitem{Lez:04}
A.{}~N.{} Leznov.
\newblock Theory of {F}ields in {Q}uantized {S}paces.
\newblock \texttt{hep-th/0409102}.

\bibitem{Luk.Now.Rue.Tol:91}
J.~Lukierski, A.~Nowicki, H.~Ruegg, and V.~N. Tolstoy.
\newblock $q$-deformation of {P}oincar\'e {A}lgebra.
\newblock {\em Phys.{} Lett.{} B}, 264:331--338, 1991.

\bibitem{Maj.Rue:94}
S.{} Majid and H.{} Ruegg.
\newblock Bicrossproduct {S}tructure of $\kappa$-{P}oincar\'e 
{G}roup and
  {N}on-commutative {G}eometry.
\newblock {\em Phys.{} Lett.{} B}, 334:348--354, 1994.

\bibitem{Vil:04}
R.~Vilela Mendes.
\newblock Some {C}onsequences of a {N}oncommutative 
{S}pace-time {S}tructure.
\newblock \texttt{hep-th/0406013}.

\bibitem{Vil:94}
R.~Vilela Mendes.
\newblock Deformations, {S}table {T}heories and {F}undamental 
{C}onstants.
\newblock {\em J.{} Phys.{} A}, 27:8091--8104, 1994.

\bibitem{Vil:00}
R.~Vilela Mendes.
\newblock Geometry, {S}tochastic {C}alculus, and {Q}uantum 
{F}ields in a
  {N}oncommutative {S}pace-time.
\newblock {\em J.{} Math.{} Phys.}, 41/1:156--186, 2000.

\bibitem{Nij.Ric:66}
A.{} Nijenhuis and R.W.{} Richardson.
\newblock Cohomology and {D}eformations in {G}raded {L}ie {A}lgebras.
\newblock {\em Bull. Amer. Math. Soc.}, 72:1--29, 1966.

\bibitem{Nij.Ric:67}
A.{} Nijenhuis and R.W.{} Richardson.
\newblock Deformations of {L}ie {A}lgebra {S}tructures.
\newblock {\em J. Math. and Mech.}, 17:89--105, 1967.

\bibitem{Ric:67}
R.W. Richardson.
\newblock On the {R}igidity of {S}emi-direct {P}roducts of 
{L}ie {A}lgebras.
\newblock {\em Pac. J. Math.}, 22:339--344, 1967.

\bibitem{Rin:79}
W.{} Rindler.
\newblock {\em Essential Relativity}.
\newblock Springer-Verlag, 1979.
\newblock (Revised Second Edition).

\bibitem{Sch.Wat.Zum:93}
P.~Schupp, P.~Watts, and B.~Zumino.
\newblock Bicovariant {Q}uantum {A}lgebras and {Q}uantum 
{L}ie {A}lgebras.
\newblock {\em Commun.{} Math.{} Phys.}, 157:305--330, 1993.
\newblock \texttt{hep-th/9210150}.

\bibitem{Shn.Ste:93}
S.{} Shnider and S.{} Sternberg.
\newblock {\em Quantum {G}roups: from {C}oalgebras to 
{D}rinfeld {A}lgebras}.
\newblock International Press, Boston, 1993.

\bibitem{Sny:47a}
H.{}~S.{} Snyder.
\newblock The {E}lectromagnetic {F}ield in {Q}uantized {S}pace-{T}ime.
\newblock {\em Phys.{} Rev.}, 72:68, 1947.

\bibitem{Sny:47}
H.{}~S.{} Snyder.
\newblock Quantized {S}pace-{T}ime.
\newblock {\em Phys.{} Rev.}, 71:38, 1947.

\bibitem{Sta:97}
J.{} Stasheff.
\newblock Deformation {T}heory and the {B}atalin-{V}ilkovisky {M}aster
  {E}quation.
\newblock In D.{} Sternheimer, J.{} Rawnsley, and S.{} Gutt, editors, 
{\em
  Deformation {T}heory and {S}ymplectic {G}eometry}. Kluwer Academic
  Publishers, Dordrecht, 1997.

\bibitem{Hol:90}
J.~W. van Holten.
\newblock The {BRST} {C}omplex and the {C}ohomology of {C}ompact {L}ie
  {A}lgebras.
\newblock {\em Nucl. Phys.}, B339:158--176, 1990.

\bibitem{Yan:47}
C.{}~N.{} Yang.
\newblock On {Q}uantized {S}pace-{T}ime.
\newblock {\em Phys.{} Rev.{}}, 72:874, 1947.

\end{thebibliography}

\end{document}